\documentclass[superscriptaddress,nofootinbib,aps,prl,floatfix,preprintnumbers,amsmath,amssymb,groupedaddress]{revtex4}

\usepackage{dsfont}
\usepackage{epsfig}
\usepackage{slashed}
\usepackage{bbold}
\usepackage{psfrag}
\usepackage{color}
\PassOptionsToPackage{caption=false}{subfig}
\usepackage{subfig}
\usepackage{multirow}
\usepackage{booktabs}
\usepackage{hyperref}
\begin{document}

\def\as{\alpha_S}
\def\t{{\bar t}}
\def\Mtt{m_{t\bar t}}
\def\PT{p_{\rm T}}
\def\PTt{p_{{\rm T},t}}
\def\GeV{\, \mathrm{GeV}}
\def\Ntot{N_{\rm tot}}
\def\None{N_{100}}
\def\Ntwo{N_{200}}

\title{Bump-hunting in LHC $t\t$ events}

\author{Michal Czakon}
\affiliation{Institut f\"ur Theoretische Teilchenphysik und Kosmologie,
RWTH Aachen University, D-52056 Aachen, Germany}

\author{David Heymes}
\affiliation{Cavendish Laboratory, University of Cambridge, Cambridge CB3 0HE, UK}

\author{Alexander Mitov}
\affiliation{Cavendish Laboratory, University of Cambridge, Cambridge CB3 0HE, UK}

\preprint{Cavendish-HEP-16/14, TTK-16-33}

\begin{abstract}
We demonstrate that a purposefully normalised NNLO $\Mtt$ differential spectrum can have very small theoretical uncertainty and, in particular, a small sensitivity to the top quark mass. Such observable can thus be a very effective bump-hunting tool for resonances decaying to $t\t$ events during LHC Run II and beyond. 
To illustrate how the approach works, we concentrate on one specific example of current interest, namely, the possible 750 GeV di-gamma excess resonance $\Phi$. Considering only theoretical uncertainties, we demonstrate that it is possible to distinguish $pp\to\Phi\to t\t$ signals studied in the recent literature [Hespel, Maltoni and Vryonidou, arXiv:1606.04149] from the pure SM background with very high significance. Alternatively, in case of non-observation, a strong upper limit on the decay rate $\Phi\to t\t$ can be placed. 
\end{abstract}
\maketitle

\section{Introduction}

{\it Bump-hunting}, i.e. searching for bumps in invariant mass spectra, is perhaps the best way to look for resonances at particle colliders. Such an approach is fairly model independent and in the limit of small bins, large statistics and resonance width that is (much) smaller than the considered kinematic range allows one to unambiguously discover and accurately map a resonance. Moreover, in this limit no detailed understanding of the relevant background is required which is a welcome feature given backgrounds are often poorly predicted.

While extremely powerful, bump-hunting search strategies have their limitations, too. The main one is limited statistics. While still an important factor in many searches, the data taking ability of the LHC will soon render statistical errors irrelevant in many cases. In the long run, especially with the high-luminosity LHC phase, statistics will become non-issue for most current searches. A second, {\it irreducible} bump-hunting limitation is finite bin size, which is introduced by statistics-independent factors like detector resolution and unfolding.
\footnote{See ref.~\cite{Spano:2013nca} for details. We thank Francesco Span\`o for discussions.}
Because of these limitations, a straightforward application of the bump-hunting approach, as described above, is not always possible. In such cases having high-precision background predictions could be very valuable.

In this work we elaborate on a search strategy for possible resonances decaying to $t\t$ final states which fully utilises the knowledge of the background with high precision ($t\t$ in this case). To make our discussion less abstract we will consider the case of the $750\GeV$ di-gamma excess \cite{ATLAS750note,CMS:2015dxe,Aaboud:2016tru,Khachatryan:2016hje}. This possible deviation from the Standard Model (SM) has triggered enormous interest and activity in direction of explaining it through beyond the SM physics (BSM); see the recent review \cite{Strumia:2016wys} for detailed cover of the existing BSM literature. We would like to stress that the approach considered in this work is general and can be adapted for different kinematics and we expect it to strengthen exclusion limits \cite{Aad:2015fna,Aad:2013nca,Aad:2012dpa} based on existing search strategies. When framed, as an example, in the context of the 750 GeV di-gamma excess, the questions we address in this work are: if the observed di-gamma excess is due to the decay of an unknown particle $\Phi\to \gamma\gamma$, could it also be observed in $t\t$ data at LHC 13 TeV? And if not observed, then how powerful a limit can be placed on the possible decay rate $\Phi\to t\t$? 

The current data suggest that the resonance $\Phi$ is most likely spin zero, has width around $40\GeV$ or less and mass around $750\GeV$ (however see sec.~{\it Note Added} about updated measurements). Model dependence aside, we assume that $\Phi$ can decay to $t\t$, which is allowed kinematically. A detailed analysis of the process $\Phi\to t\t$ has been performed recently in refs.~\cite{Hespel:2016qaf,Djouadi:2016ack}; similar analysis, not directly related to the $750\GeV$ di-gamma excess, has also been performed in refs.~\cite{Bernreuther:2015fts,Craig:2015jba,Gori:2016zto}. Ref.~\cite{Hespel:2016qaf} presents predictions for a number of models with production rates $\sigma(pp\to\Phi\to t\t)$ between 0.2 pb and 1.2 pb. In the following, we utilise the predictions of ref.~\cite{Hespel:2016qaf} 
\footnote{We thank the authors of ref.~\cite{Hespel:2016qaf} for kindly providing us with their results in electronic form.}
and combine them with the recent NNLO QCD calculation of the $\Mtt$ spectrum \cite{Czakon:2015owf,Czakon:2016dgf} to demonstrate the potential for discriminating BSM models for $\Phi\to t\t$ from the SM $t\t$ background in the LHC $\Mtt$ spectrum.

\section{The search strategy}

\begin{figure}[t]
\centering
\hspace{0mm} 
\includegraphics[width=0.32\textwidth]{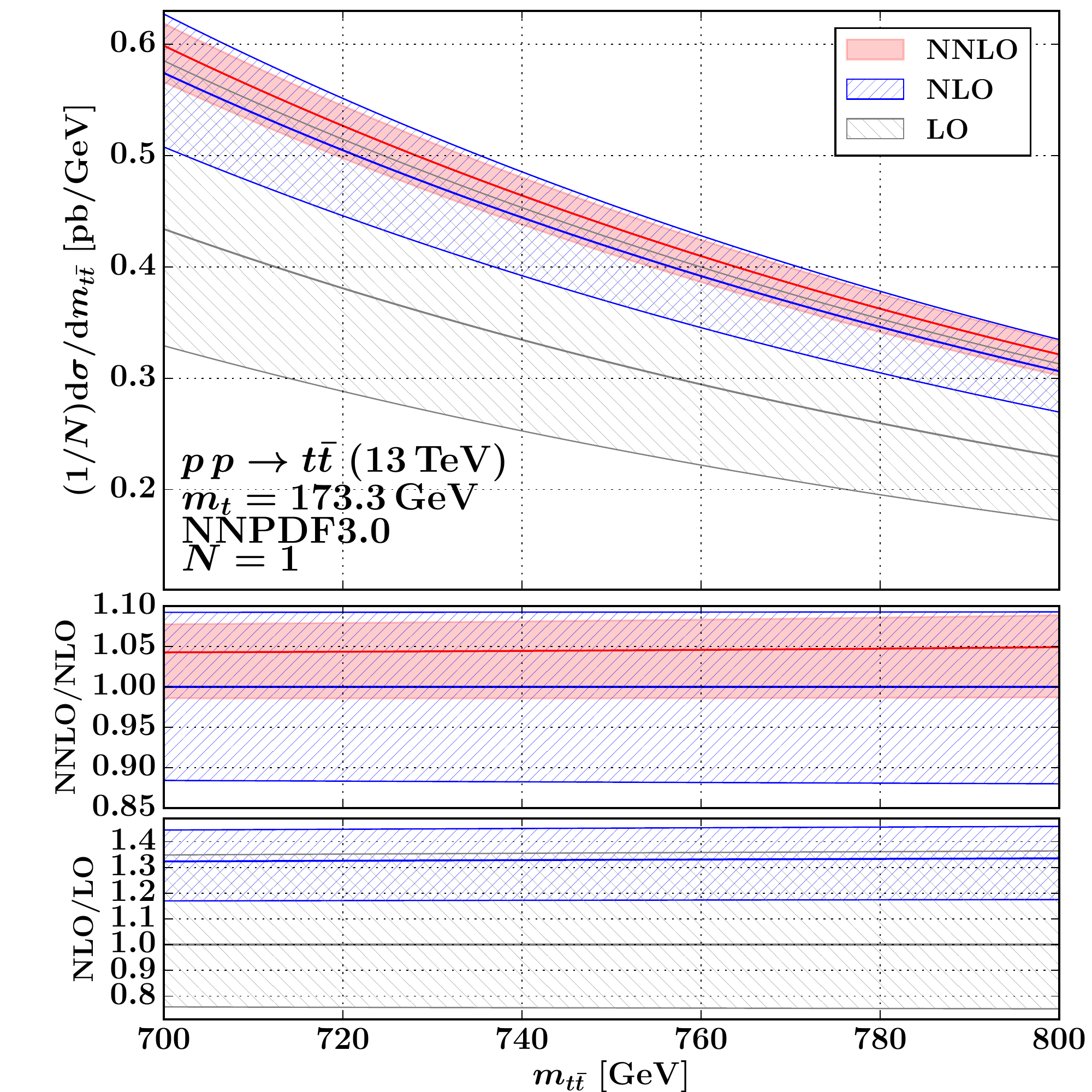}
\includegraphics[width=0.32\textwidth]{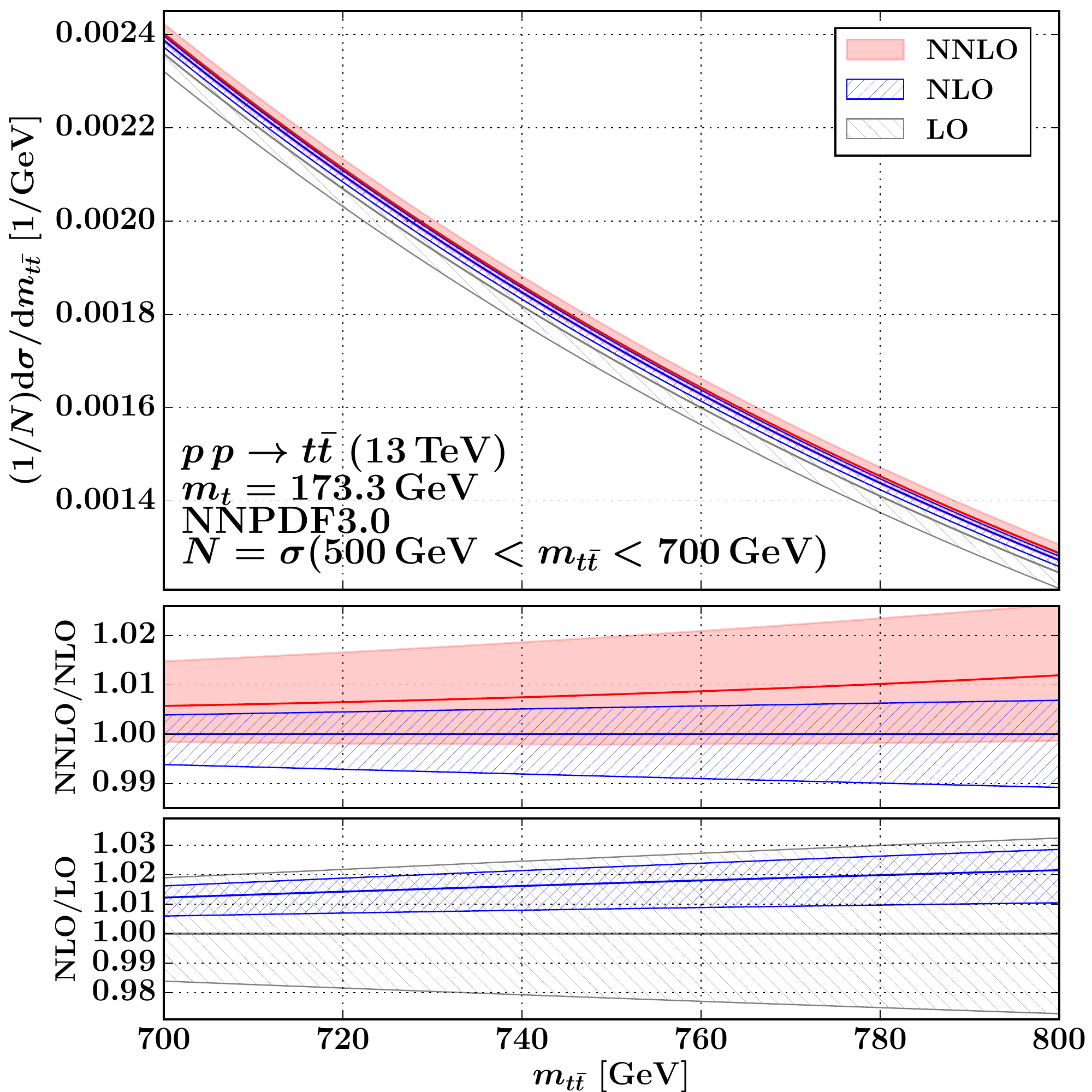}
\includegraphics[width=0.32\textwidth]{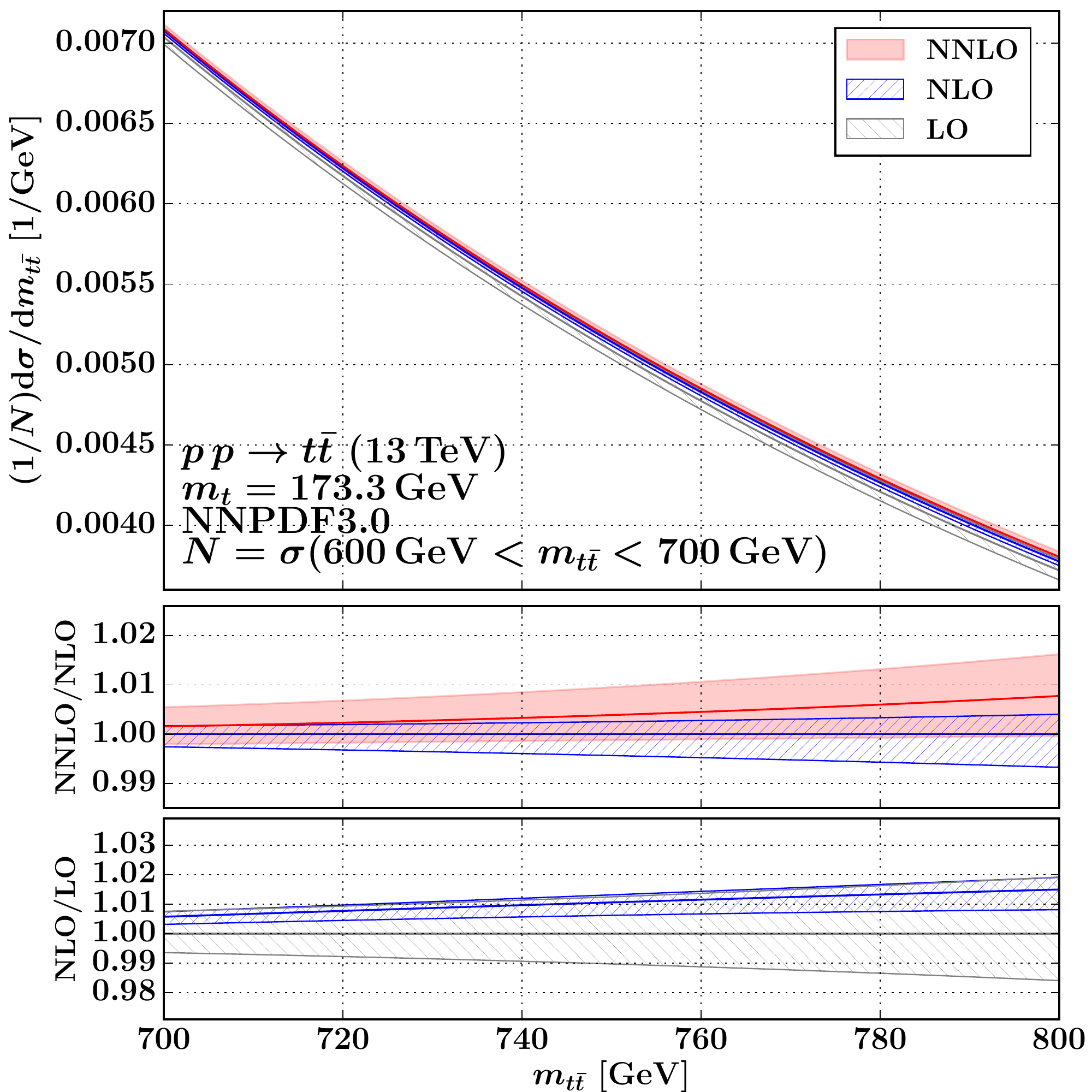}
\caption{The $\Mtt$ distribution in the range $(700,800)\GeV$ through NNLO QCD: with absolute normalisation $N=1$ (left), normalised to $N_{200}=\sigma(500\GeV<\Mtt <700\GeV)$ (centre) and normalised to $N_{100}=\sigma(600\GeV<\Mtt <700\GeV)$ (right). The normalisation factor $N$ is introduced in eq.~(\ref{eq:Norm-dist}) below. The error band at NNLO is from pdf and scale variation added in quadrature. Only the scale error is shown at LO and NLO.}
\label{fig:Mtt-abs}
\end{figure}

The simplest search strategy is to look for bumps in the unnormalised $\Mtt$ spectrum. In fig.~\ref{fig:Mtt-abs}  we show the $\Mtt$ spectrum in LO, NLO and NNLO QCD. We use the dynamic scale results of ref.~\cite{Czakon:2016dgf}. All pure QCD calculations are done with the NNPDF3.0 pdf set \cite{Ball:2014uwa}. We use $m_t=173.3\GeV$ and estimate scale error, as usual, through independent factorisation and renormalisation scale variation \cite{Cacciari:2008zb}. The total error at NNLO is obtained by adding pdf and scale errors in quadrature. At LO and NLO we show only the scale errors. To compute the pdf error at NNLO we use the rescaling approximation detailed in sec.~3.1 of ref.~\cite{Czakon:2016ckf}. The findings of ref.~\cite{Czakon:2016dgf} show that this procedure should work very well in the present LHC context, too. 

In view of the application to the possible $750\GeV$ di-gamma excess, motivated in the Introduction, henceforth we restrict our discussion of the $t\t$ invariant mass distribution to the interval $700\GeV <\Mtt <800\GeV$. This specification aside, our discussion remains fully general.

The differential distribution showed in fig.~\ref{fig:Mtt-abs} is, in fact, an analytic fit to the finest binned differential $\Mtt$ distribution computed in ref.~\cite{Czakon:2016dgf}. Given the smoothness of the differential distribution in this $\Mtt$ range, as well as the small MC error of the underlying binned calculation (see ref.~\cite{Czakon:2016dgf} for details) it is natural to derive such a fit. Most importantly, an analytic fit allows one to subsequently derive binned distributions with any bin size, or bin position, and we will explore both possibilities in the following. We fit separately the central prediction as well as the lower and upper edges in each bin. We repeat this procedure for LO, NLO and NNLO bands in the cases of absolute normalisation as well as the two normalisations to be introduced shortly. To ensure smoothness of the fit, we perform it over the wider range $600\GeV <\Mtt <900\GeV$ although the fit is only meant to be used in the range $700\GeV <\Mtt <800\GeV$. 

In all cases the fits take the functional form:
\begin{equation}
{d\sigma\over d\Mtt} = c_1+c_2\,e^{- c_3 \Mtt}\,.
\label{eq:fit}
\end{equation}
The fit coefficients $c_{1,2,3}$ are available in electronic form with the Arxiv submission of this paper. The quality of the fits is such that the relative scatter of the actual calculation with respect to its fit is between 0.2\% and 0.7\% for all bins in the interval $700\GeV <\Mtt <800\GeV$. Moreover, the scatter is consistent with both being random and with the estimated MC error.

In the following we detail the most important sources of theory error and how we deal with them. 

Scale error is estimated through independent scale variation, as usual. We believe that in the $700\GeV <\Mtt <800\GeV$ range scale variation is a good estimator of missing higher order effects, because in this intermediate $\Mtt$ range neither absolute threshold Coulomb effects nor collinear log resummation play a role. Soft-gluon resummation might have an effect but following the findings of ref.~\cite{Czakon:2016dgf} we expect that resummation effects, when properly matched to the NNLO calculation used in the present work, should be within the NNLO scale error estimate. 

Pdf error should also be under good control in this $\Mtt$ range. A detailed analysis \cite{Czakon:2016olj} shows that there is a good overlap between various state-of-the-art pdf sets. By taking NNPDF3.0 as our default set we likely have a conservative estimate of the pdf error.

Electroweak (EW) corrections contribute little in this $\Mtt$ range. Utilising the recent work \cite{Pagani:2016caq}, a detailed analysis of mixed NNLO QCD and EW corrections in top production \cite{NNLO-EW} shows that the $\Mtt$ distribution is slightly lowered by EW corrections (by about 1-2\%). Such an effect is negligible in the unnormalised $\Mtt$ distribution. We have checked that EW effects are reduced to below 0.3\% in the normalised distributions to be defined next, and so we neglect them in the following.

Another subtle source of theory error is the value of the top quark mass (throughout this work we utilise the top quark pole mass). With a direct calculation at LO and NLO
\footnote{We would like to thank Michelangelo Mangano for a useful suggestion.}
we estimate that a $1\GeV$ change in $m_t$ (with respect to $m_t=173.3\GeV$) shifts the differential cross-section around $\Mtt=750\GeV$ by about 1\%; see Appendix for details. Since the error of the current $m_t$ world average is well below $1\GeV$ \cite{ATLAS:2014wva} it may appear that the $m_t$ systematics is not important. There are two indications, however, that this may not be the case. First, there is a spread of around $3\GeV$ between independent precise measurements of $m_t$ \cite{CMS:2014ima,Abazov:2014dpa} (a recent summary of LHC measurements can be found in ref.~\cite{Stieger:2016vgj}) which, when coupled with the discussion of ref.~\cite{Frixione:2014ala}, indicates that robust control over the $m_t$ systematics is prudent in the present context. Second, the $m_t$ systematics may play an outsized role in the normalised $\Mtt$ distribution. To aid the following discussion we introduce the normalised $\Mtt$ differential distribution parametrised by the normalisation factor $N$:
\begin{equation}
\sigma(N) = {1\over N}{d\sigma\over d\Mtt} \,.
\label{eq:Norm-dist}
\end{equation}
Of interest to us will be the following normalisation factors:
\begin{eqnarray}
N&=&1 ~~({\rm i.e.~the~unnormalised~distribution})\,,\nonumber \\
N_{\mathrm{tot}}&=&\sigma_{\mathrm{tot}}\,,\nonumber\\
N_{100}&=&\sigma(600\GeV< \Mtt <700\GeV)\,,\nonumber\\
N_{200}&=&\sigma(500\GeV< \Mtt <700\GeV)\,.
\label{eq:norm}
\end{eqnarray}

\begin{figure}[t]
\centering
\hspace{0mm} 
\includegraphics[width=0.47\textwidth]{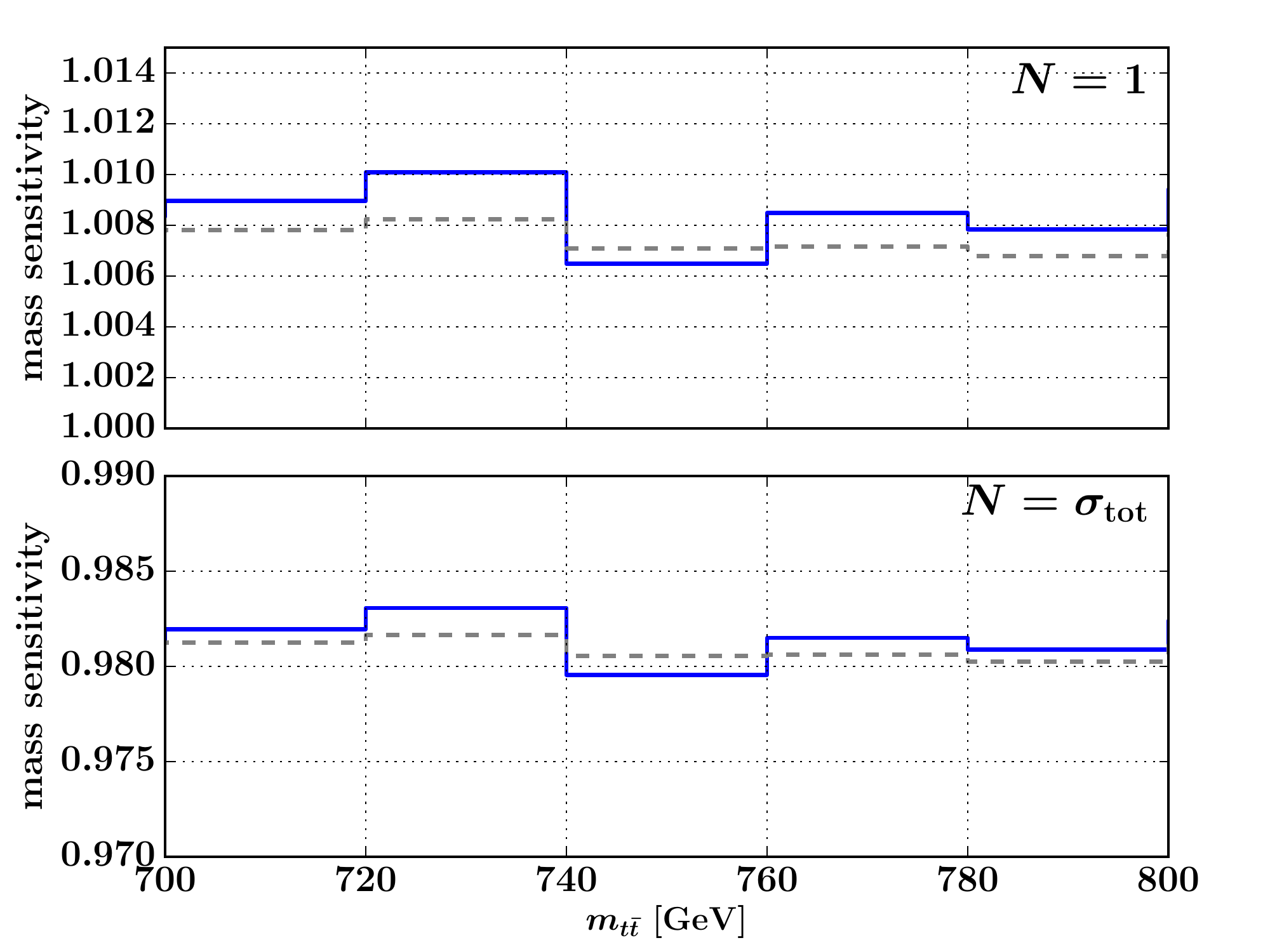}
\includegraphics[width=0.47\textwidth]{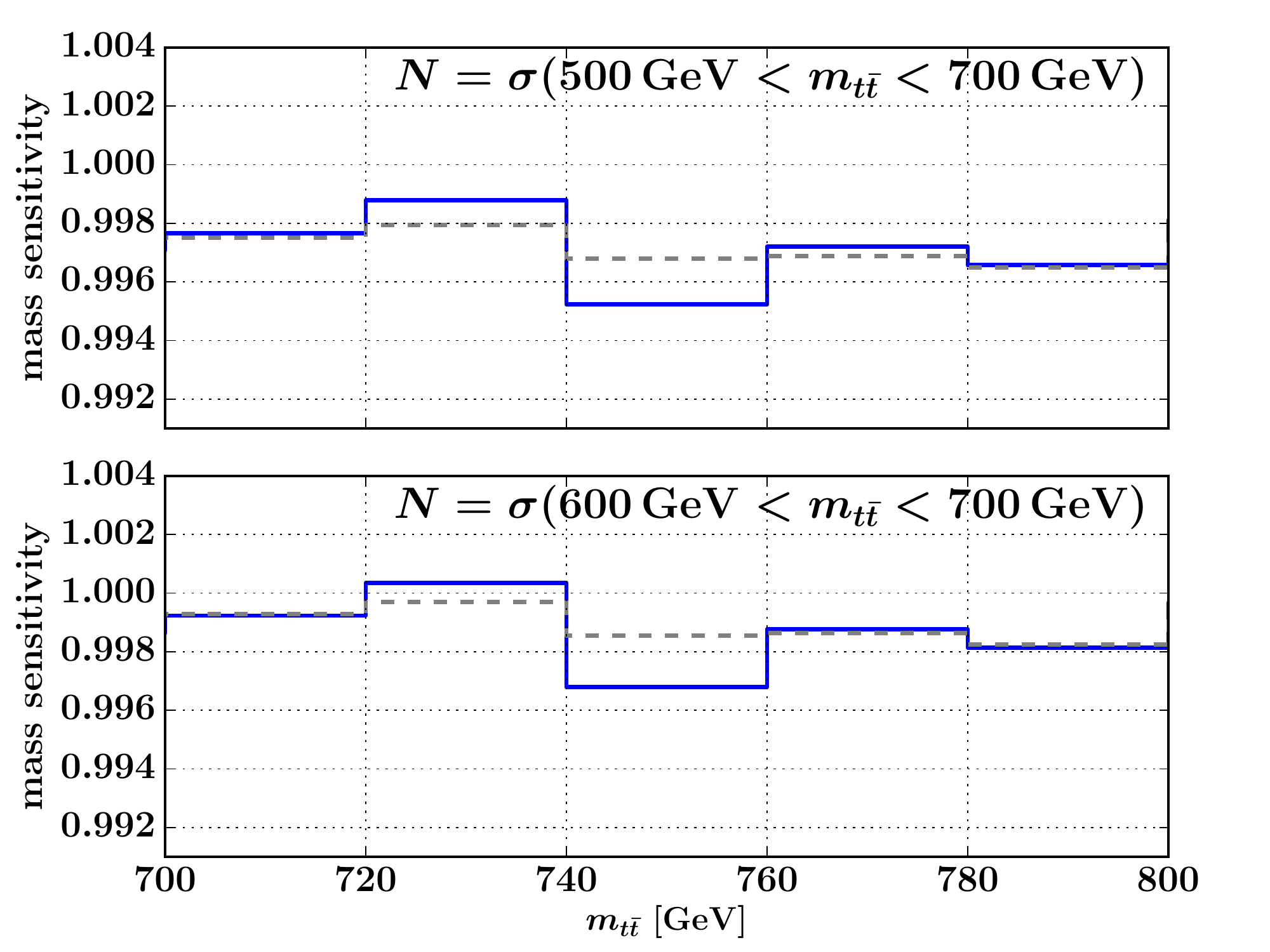}
\caption{The $m_t$ sensitivity of the differential $\Mtt$ distribution around $\Mtt=750\GeV$ for the four normalisations eq.~(\ref{eq:norm}).}
\label{fig:mass-ratio}
\end{figure}

A quick check, see fig.~\ref{fig:mass-ratio}, shows that in the $\Mtt$ range around $\Mtt=750\GeV$ the usual normalised distribution $\sigma(\Ntot)$ has twice the $m_t$ sensitivity of the unnormalised distribution $\sigma(1)$. The differential sensitivity is defined as:
\begin{equation}
\mathrm{mass}\,\, \mathrm{sensitivity} = \frac{d\sigma(m_t=172.3\GeV)}{d\sigma(m_t=173.3\GeV)}\,.
\label{eq:mass-sensitivity}
\end{equation}

Since both scale and pdf errors get strongly reduced in normalised distributions, see below, the $m_t$ sensitivity may turn out to be a leading theoretical systematics for $\sigma(\Ntot)$ in this $\Mtt$ range. For this reason we will not consider the $\sigma(\Ntot)$ distribution in this work.

It turns out that in order to minimise the $m_t$ sensitivity in the relevant for this work portion of the $\Mtt$ spectrum, we need to normalise the differential spectrum to the inclusive cross-section based on $\Mtt$ values just below the range we are interested in. To be specific, in this work we consider the two normalisations $\None$ and $\Ntwo$ defined in eq.~(\ref{eq:norm}). The reasons behind choosing their ranges are as follows. We consider as upper limit $700\GeV$ because this is the lower end of the window $700\GeV<\Mtt <800\GeV$ where we intend to search for $\Phi\to t\t$ decays. Assuming $m_\Phi\approx 750\GeV$ and $\Gamma_{\Phi\to t\t}<40\GeV$ we do not expect to have much pure BSM signal below $\Mtt=700\GeV$ (the interference with the SM $t\t\,$ background, however, does contribute - see below). There are two competing demands when trying to decide on the  size of the normalisation window. A larger normalisation interval would, presumably, minimise experimental errors. On the other hand, a smaller normalisation interval will lead to smaller sensitivity to $m_t$. Indeed, with direct LO and NLO calculations, we have estimated that the $m_t$ sensitivity of the distributions $\sigma(\None)$ and $\sigma(\Ntwo)$ is much reduced in the interval around $\Mtt=750\GeV$: as can be seen from fig.~\ref{fig:mass-ratio}, it is below $0.2\%$ per $\GeV$ for $\sigma(\None)$ and around $0.3\%$ per $\GeV$ for $\sigma(\Ntwo)$. Additionally, the error due to $\as$ should also be strongly reduced in these two normalised distributions compared to the unnormalised one. In this work we have not further investigated the sensitivity to $\as$.

Next we address the properties of the normalised distributions eq.~(\ref{eq:Norm-dist}) shown in fig.~\ref{fig:Mtt-abs}. The main feature of the normalised distributions is their strongly reduced scale and pdf variation. Indeed, as can be seen from fig.~\ref{fig:Mtt-abs} the combined scale and pdf error for the case $N_{200}$ is about 1\% while for $N_{100}$ it is only around half that, i.e. 0.5\%. As it is often the case in such normalised observables, such strong reduction in the error estimate is driven by the fact that scale and pdf variations are performed in a correlated way for the numerator and denominator in the normalised distribution. In other words in order to determine the scale variation one computes the numerator and denominator for the same choice of scales and then studies the variation of the ratio. Similarly for the pdf error. 

One may wonder if such a strong reduction in scale and pdf variation, which is the result of a consistent theoretical calculation, properly reflects the error on the ratio itself. Past examples, notably the top-pair forward-backward asymmetry through NNLO \cite{Czakon:2014xsa,Czakon:2016ckf}, show that error estimates of ratios may be more delicate than for standard observables and one should be alerted to the possibility for underestimating theoretical errors in ratios.

While in general we share such concerns, in this specific case we anticipate that the scale and pdf error estimates given above are reliable. This can be justified with the  $K$-factors shown in fig.~\ref{fig:Mtt-abs}. Unlike the unnormalised distribution $\sigma(1)$ which has large NLO and NNLO $K$-factors, the normalised distributions $\sigma(\None)$ and $\sigma(\Ntwo)$ have extremely small $K$-factors, at or below 1\%, for both NLO and NNLO. Furthermore, as also evident from fig.~\ref{fig:Mtt-abs}, the scale plus pdf error is consistent with the NLO and NNLO $K$-factors.

\section{Adding the $\Phi\to t\t$ signal}

As we mentioned above, we utilise the $\Phi\to t\t$ signal as well as the signal-background interference as calculated in ref.~\cite{Hespel:2016qaf}. Specifically, we take the model specified in table 9 of ref.~\cite{Hespel:2016qaf} which corresponds to a scalar $\Phi$ with production cross-section $\sigma(pp\to \Phi\to t\t) = 1.1\,{\rm pb}$. We combine our calculation of the SM QCD $t\t$ background with the BSM signal by simply adding the non-SM contributions computed in ref.~\cite{Hespel:2016qaf} (pure signal plus signal-background interference) to the pure SM QCD background. While this procedure is formally correct, some small inconsistencies are present. For example, ref.~\cite{Hespel:2016qaf} uses $m_t=173\GeV$ (we use $m_t=173.3\GeV$); LO calculation of the SM background with dynamic scale $\mu=\Mtt/2$ (we use scale $\mu=H_T/4$; see ref.~\cite{Czakon:2016dgf} for details); pdf set MMHT2014 \cite{Harland-Lang:2014zoa} (we use NNPDF3.0). Given the exploratory nature of this work, however, such inconsistencies are unlikely to play a role into the conclusions drawn in the following.

\begin{figure}[h]
\centering
\hspace{0mm} 
\includegraphics[width=0.32\textwidth]{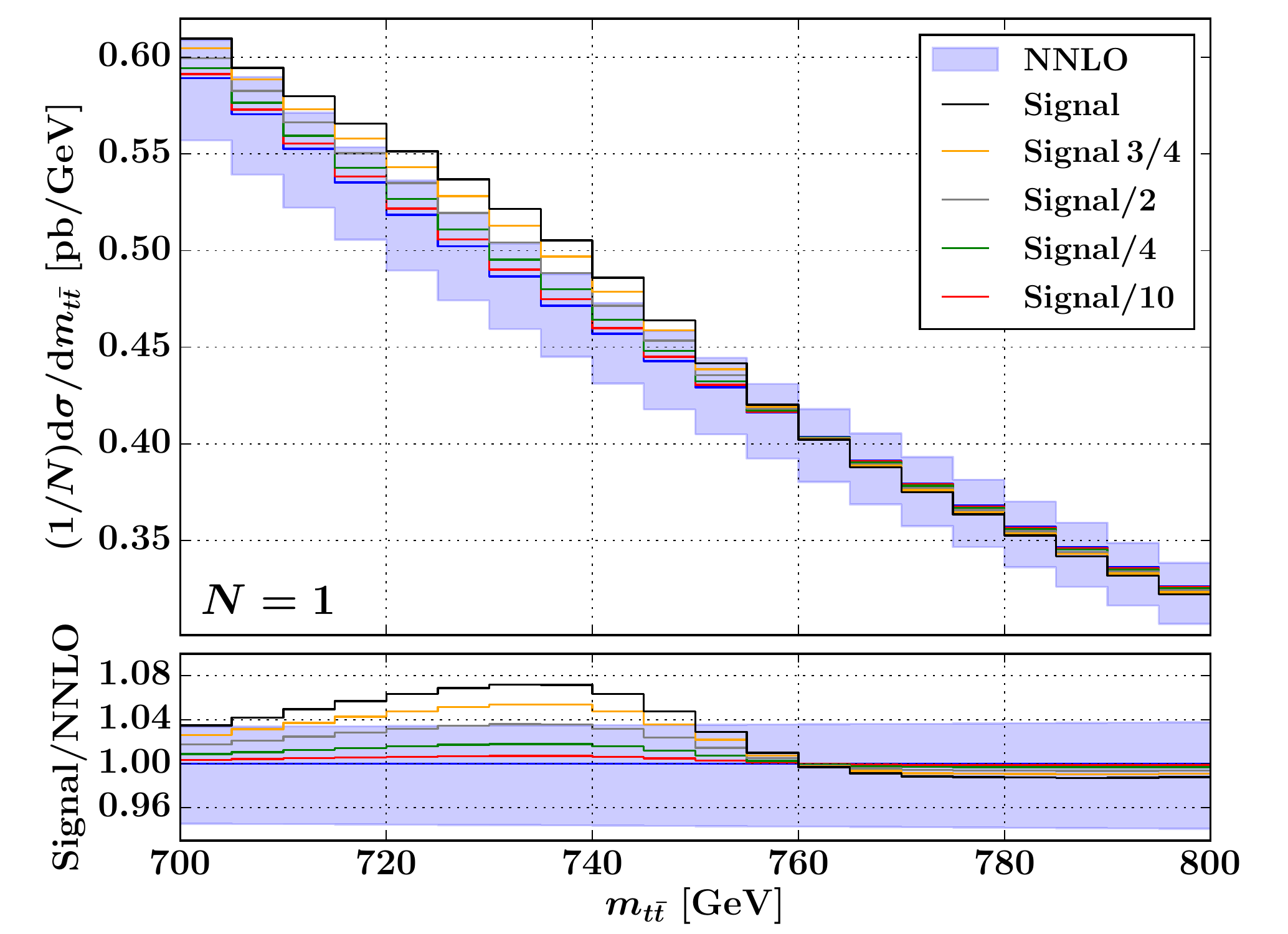}
\includegraphics[width=0.32\textwidth]{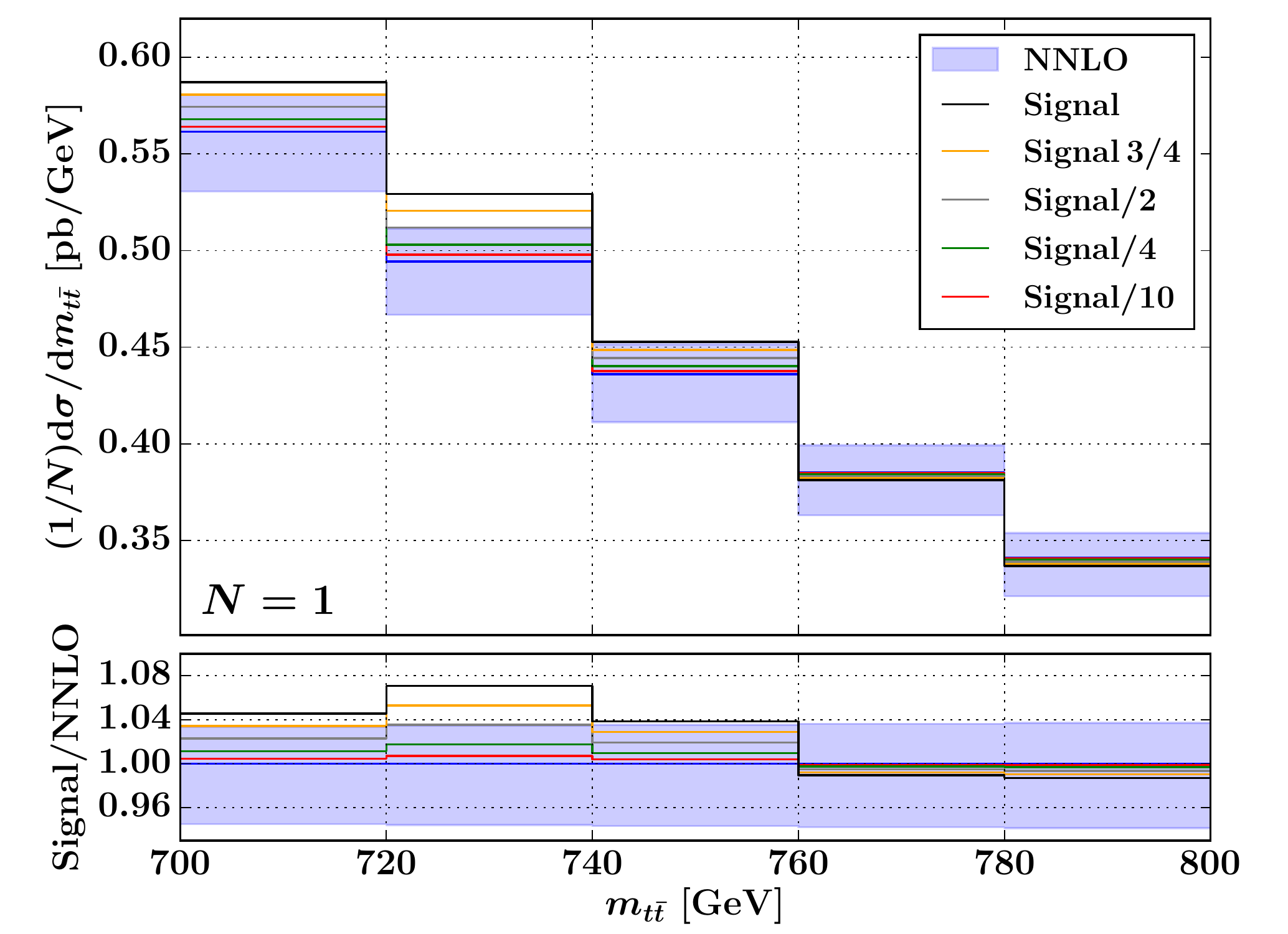}
\includegraphics[width=0.32\textwidth]{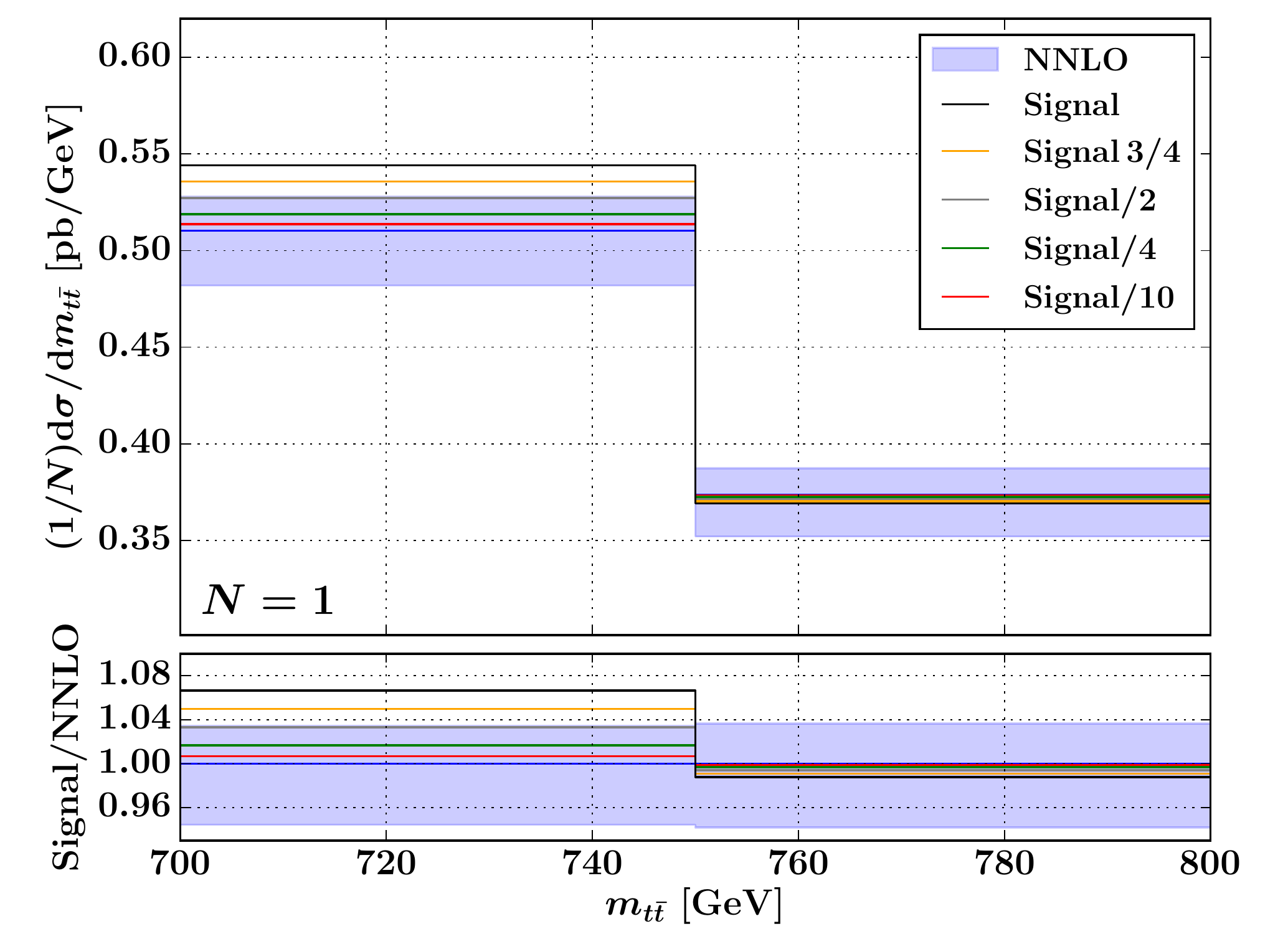}
\caption{The unnormalised $\Mtt$ distribution computed in NNLO QCD (blue band) versus NNLO QCD plus the $\Phi\to t\t$ contribution (``Signal"). Also shown are contributions from fake BSM signals derived from the nominal one, ``Signal", by simply dividing  its contribution to each bin with the indicated factor. Three different bin sizes are used: $5\GeV$ (left), $20\GeV$ (centre) and $50\GeV$ (right).}
\label{fig:Mtt-abs-binned}
\end{figure}
\begin{figure}[h]
\centering
\hspace{0mm} 
\includegraphics[width=0.32\textwidth]{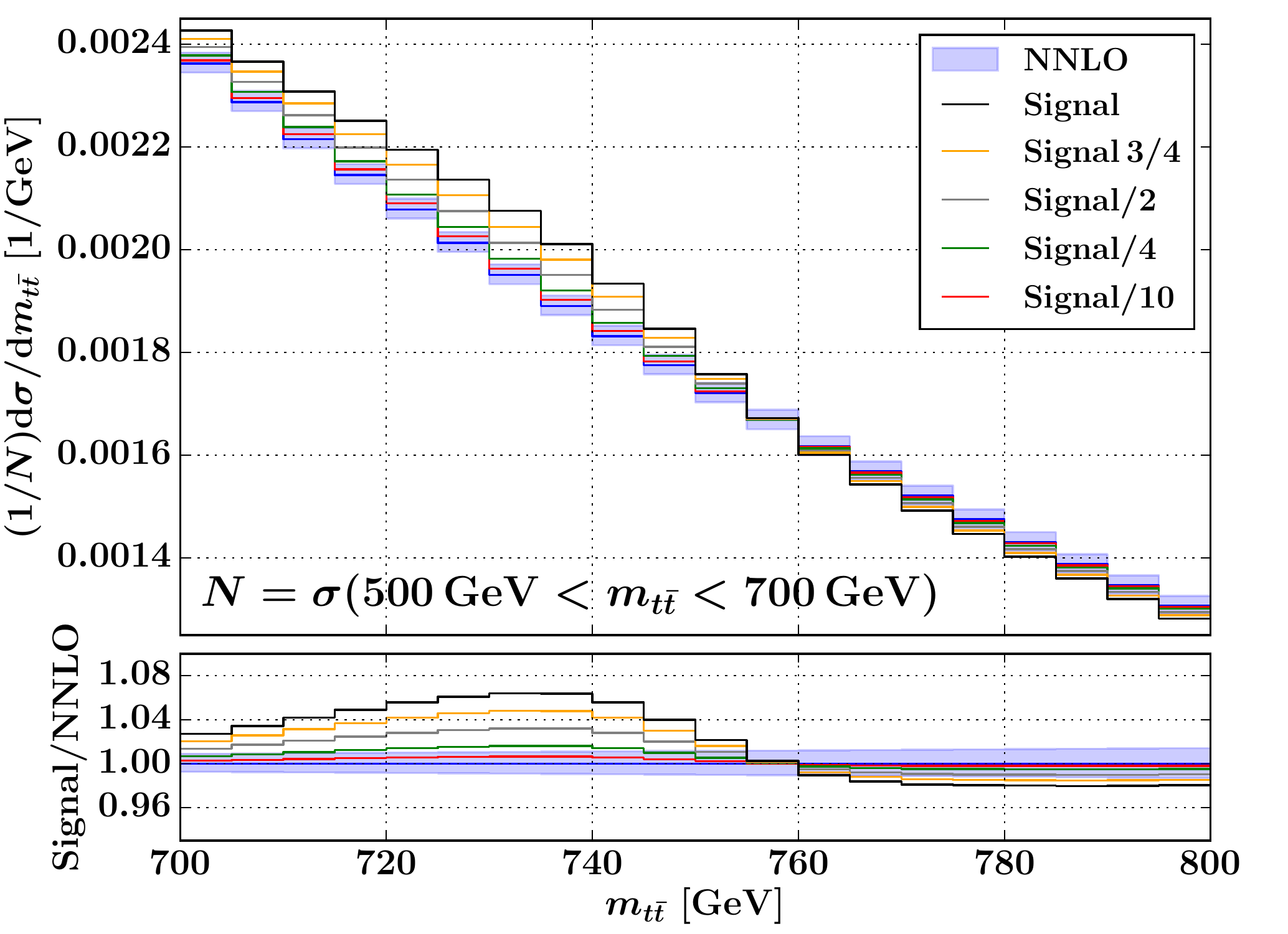}
\includegraphics[width=0.32\textwidth]{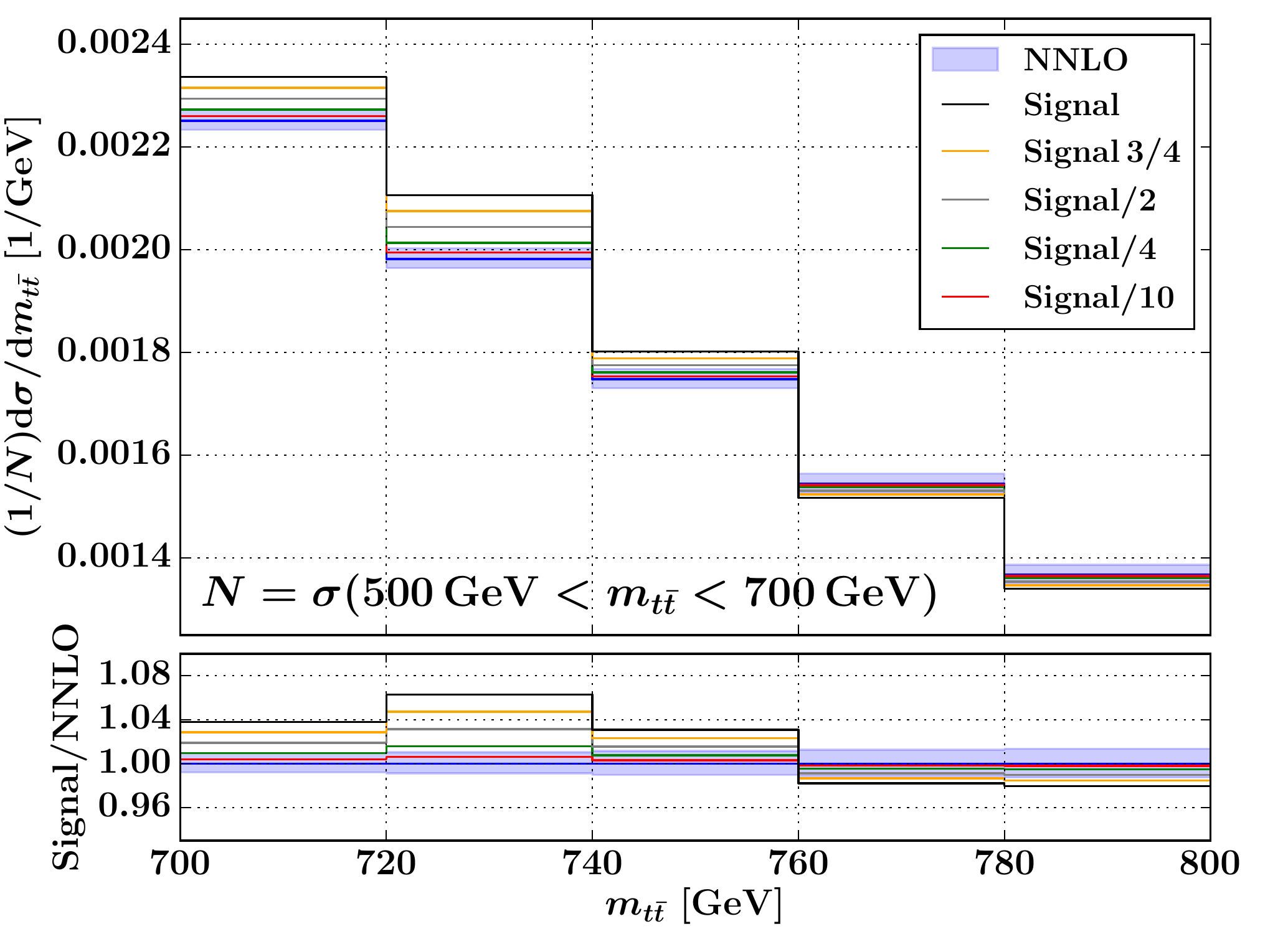}
\includegraphics[width=0.32\textwidth]{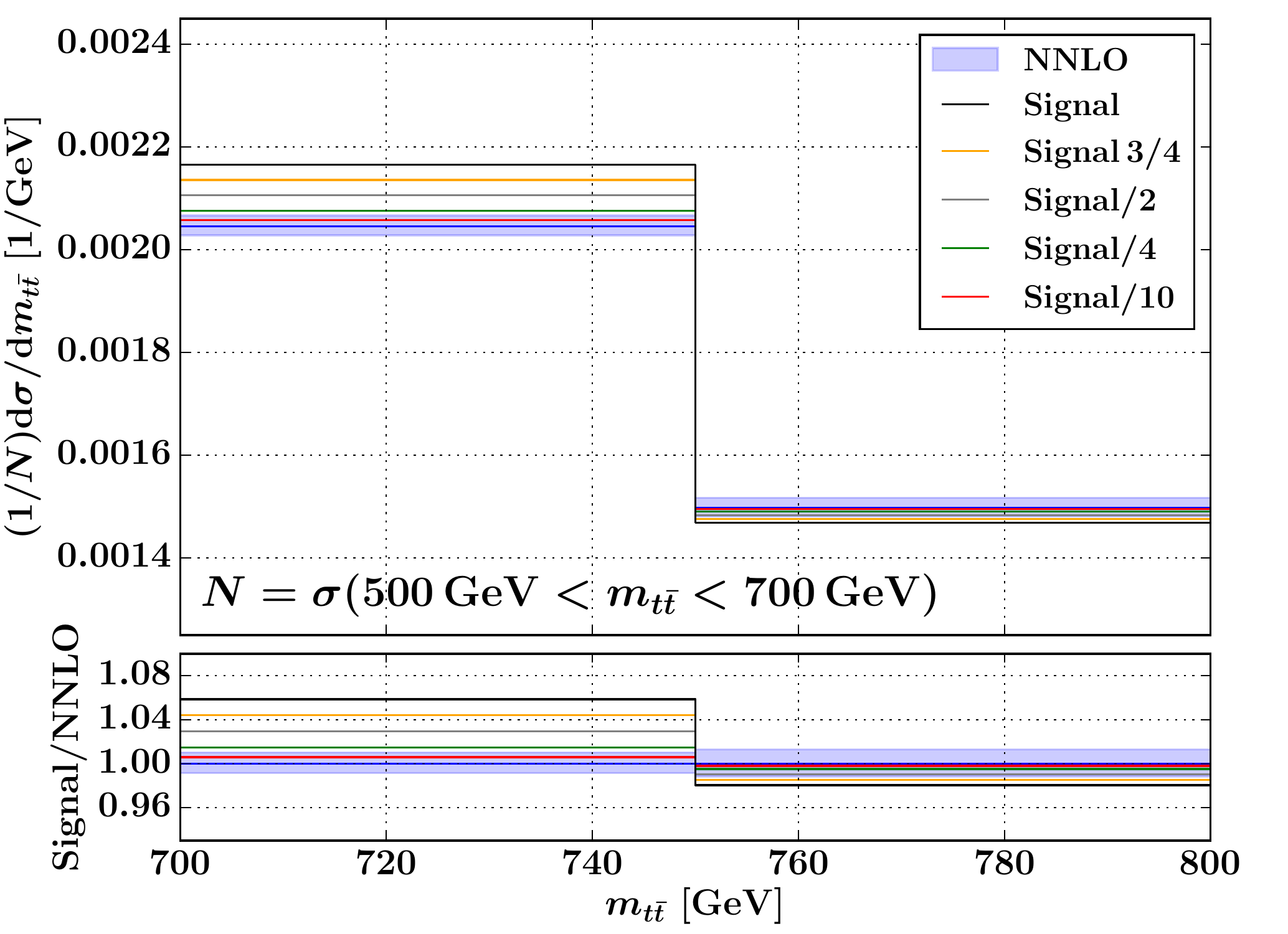}
\includegraphics[width=0.32\textwidth]{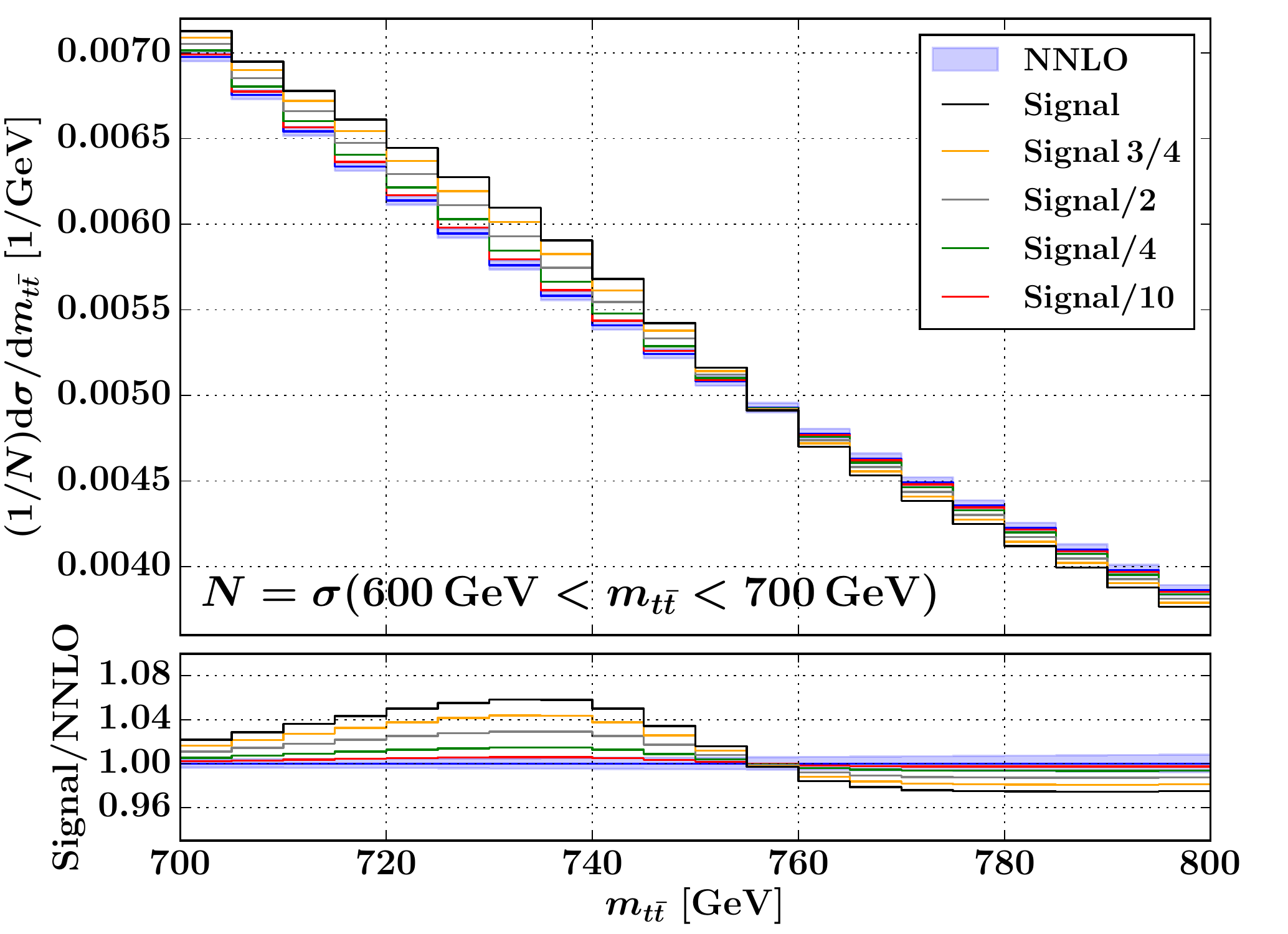}
\includegraphics[width=0.32\textwidth]{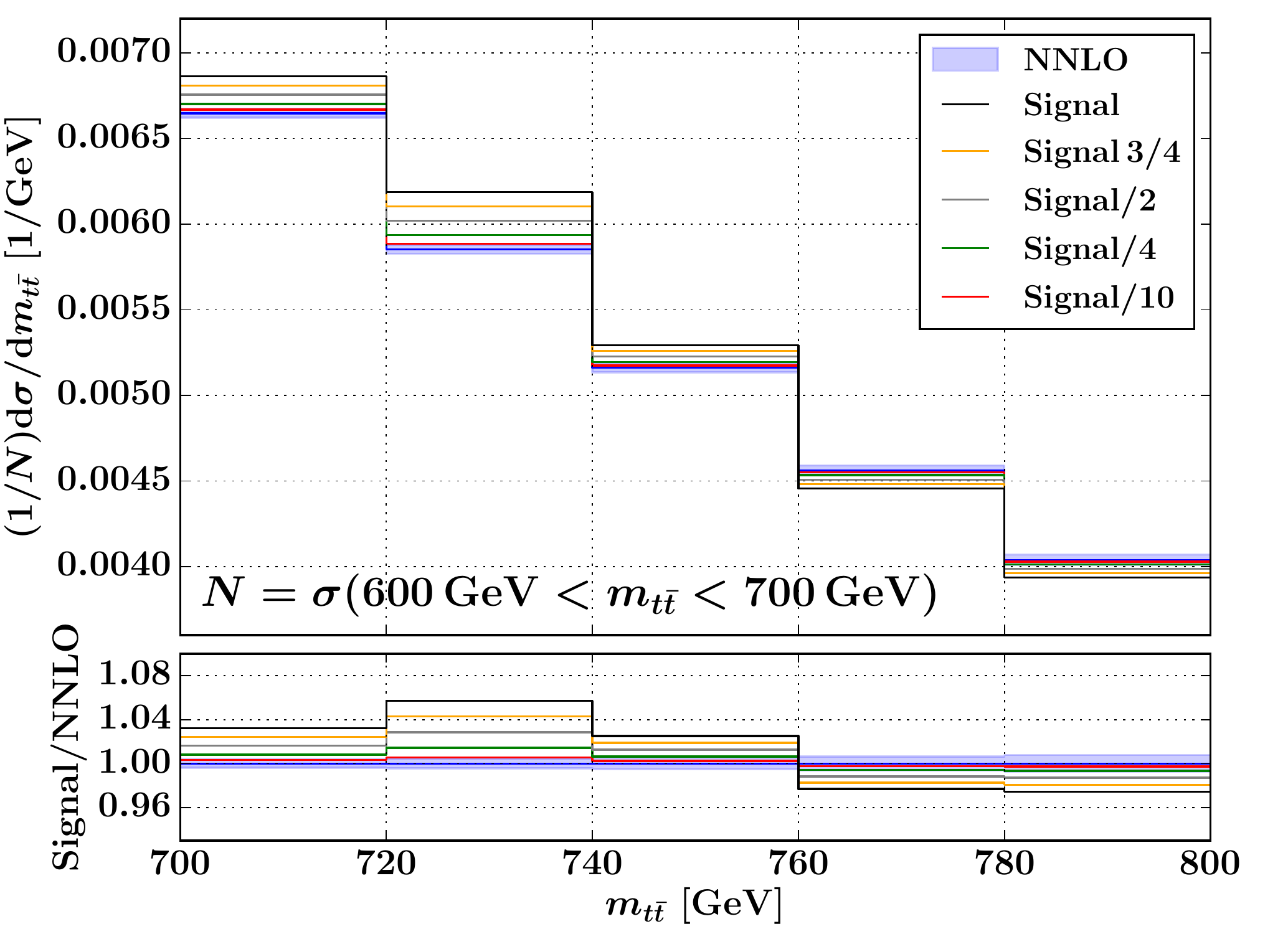}
\includegraphics[width=0.32\textwidth]{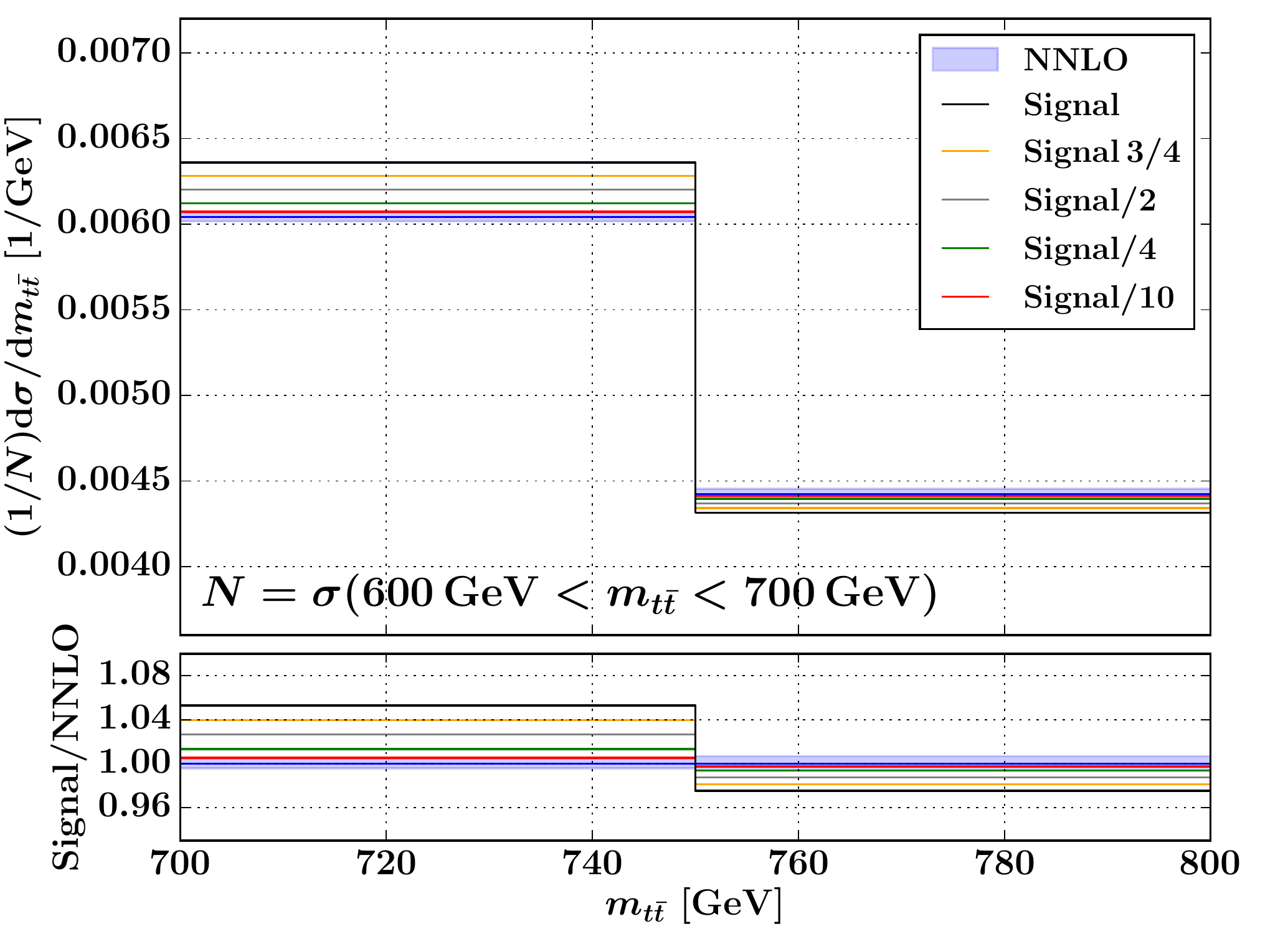}
\caption{As in fig. \ref{fig:Mtt-abs-binned} but for the normalised $\Mtt$ distributions $\sigma(\Ntwo)$ (top row) and $\sigma(\None)$ (bottom row).}
\label{fig:Mtt-norm-binned}
\end{figure}

In fig.~\ref{fig:Mtt-abs-binned} we show the complete SM+BSM contribution to the unnormalised $\Mtt$ spectrum. In order to illustrate the discriminating power of the approach we also show the cases of fake BSM signals that are derived from the BSM model of ref.~\cite{Hespel:2016qaf} considered in this work, by dividing its contribution in each bin by a constant factor. We consider bins of three sizes: $5,20$ and $50\GeV$. Of relevance for experimental analyses is only the $50\GeV$ bin size since, to our knowledge, this is the minimum bin size that will be possible in this $\Mtt$ range due to resolution constraints. Nevertheless, smaller bin sizes provide insight into the precise behaviour of signal and background. The same plots but for the normalised $\Mtt$ distributions are shown in fig.~\ref{fig:Mtt-norm-binned}: for $\sigma(\Ntwo)$ (upper row) and for $\sigma(\None)$ (lower row). In order to be able to study variable bin sizes and positions, we have derived analytical fits for the pure BSM signal and SM-BSM interference which were computed for fixed bin sizes in ref.~\cite{Hespel:2016qaf}.

The contribution of the interference between BSM signal and SM background to the normalisation factors $N_{100,200}$ is around 1\% and is included in the normalisation factor for the SM+BSM case (but, of course, it is not in the calculation of the pure SM background and normalisation). The Monte Carlo error of the normalisation factors is of particular concern since it shift up/down the whole distribution and such an error cannot be detected by the usual smoothness requirement (which only helps identify bin-to-bin MC fluctuations). We estimate the MC error of the normalisation factors directly form our calculation and find it to be below 0.2\%. Such error is insignificant and we will neglect it in the following.

In all plots in figs.~\ref{fig:Mtt-abs-binned},\ref{fig:Mtt-norm-binned} the blue bands represent the combined scale plus pdf error. Owing to the much reduced error of the normalised distributions the significance of the deviation of the signal plus background with respect to pure background is much larger. It allows to effectively distinguish not only the specific $\Phi$ model considered here but also models that predict significantly smaller value for the total rate $\sigma(pp\to \Phi\to t\t)$. 
\begin{figure}[b]
\centering
\hspace{0mm} 
\includegraphics[width=0.32\textwidth]{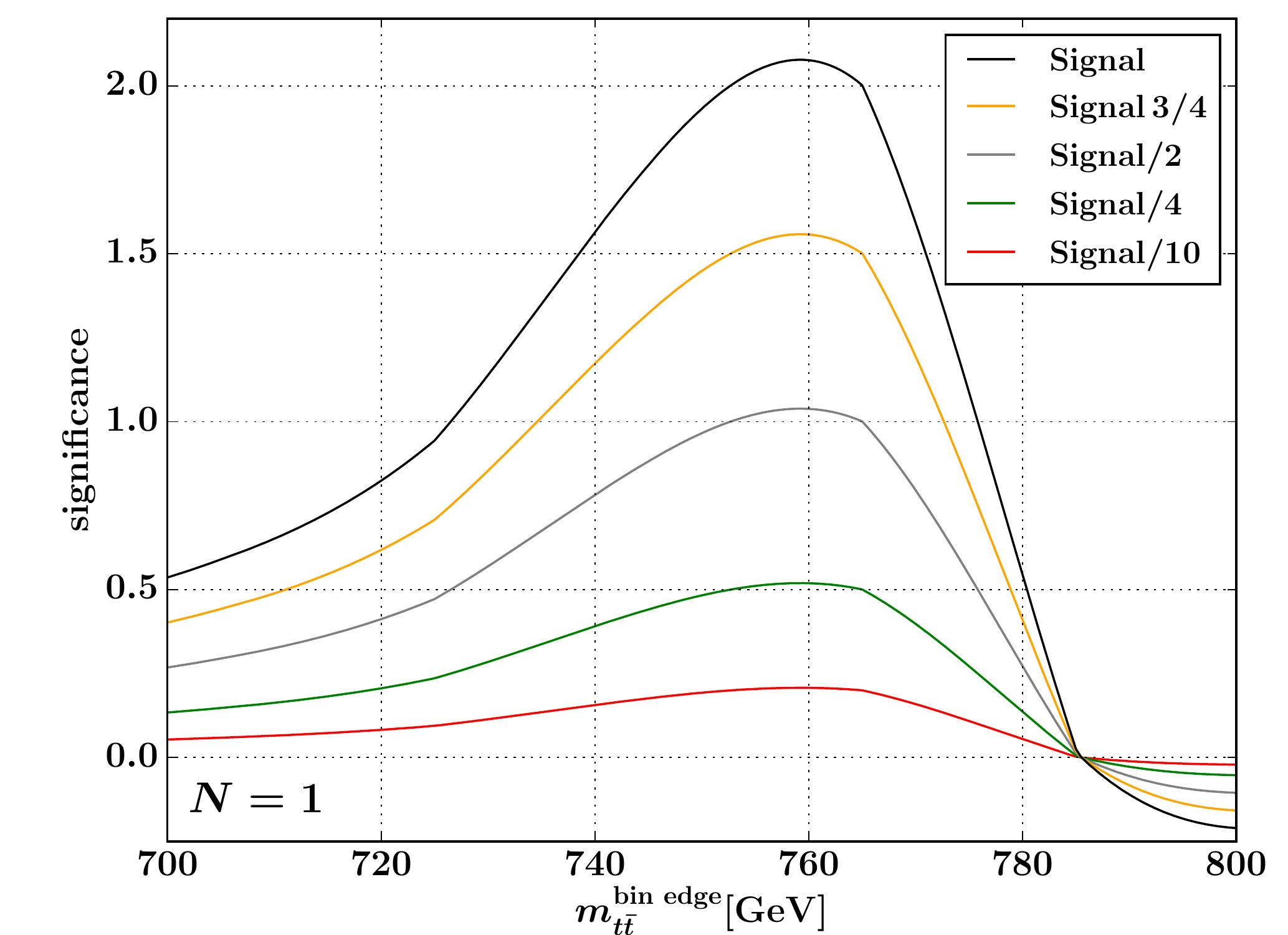}
\includegraphics[width=0.32\textwidth]{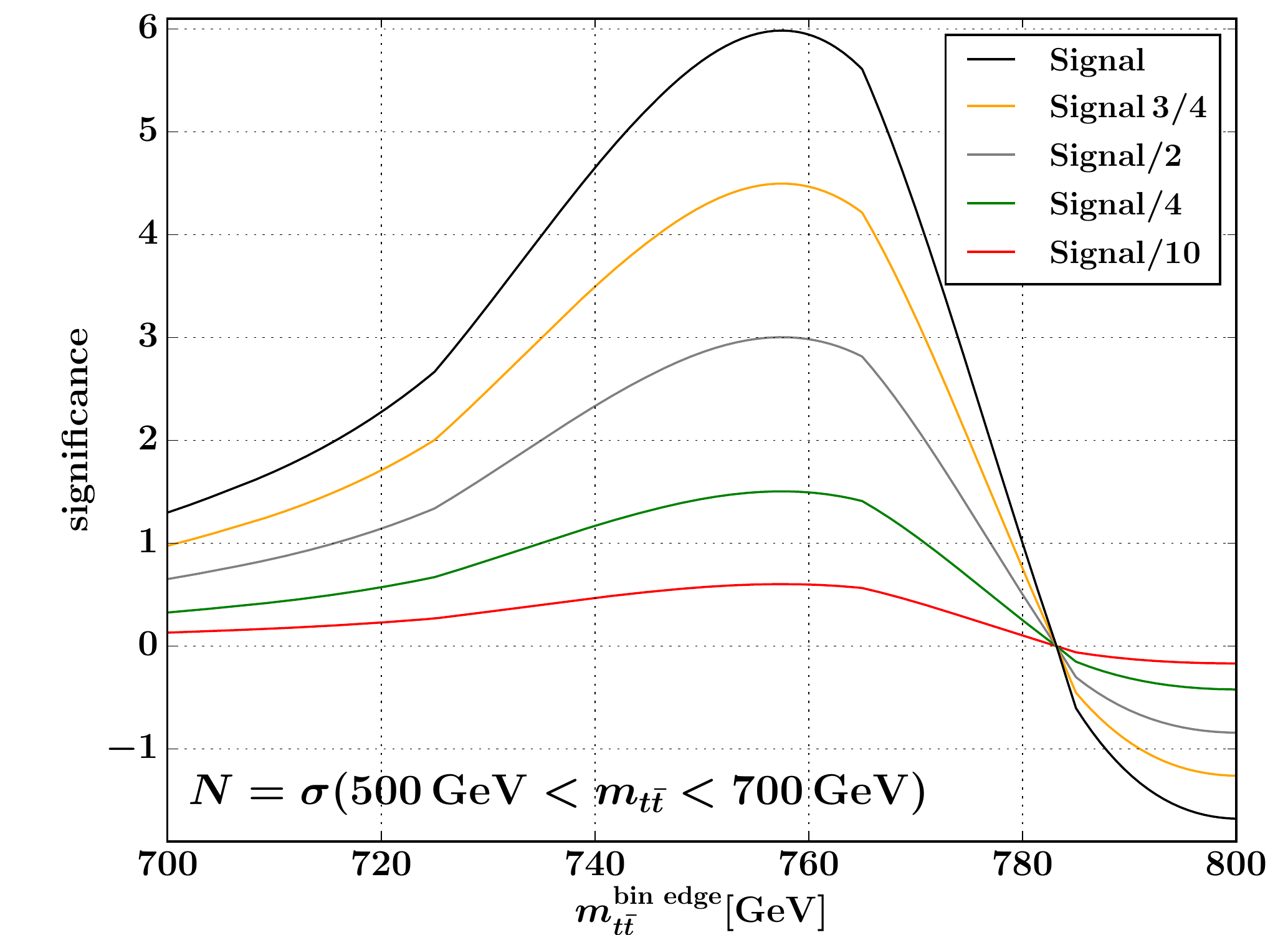}
\includegraphics[width=0.32\textwidth]{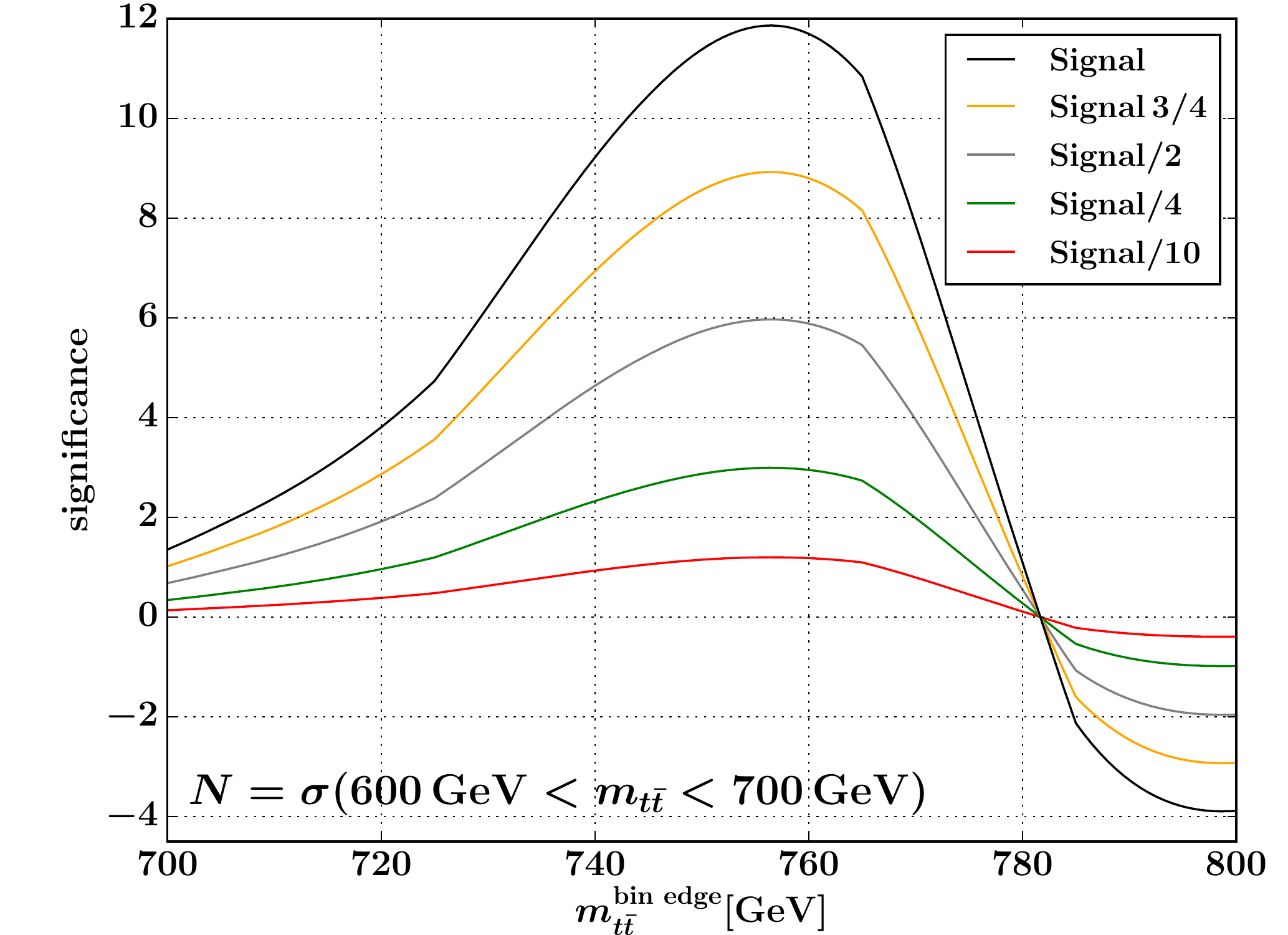}
\caption{Significance eq.~(\ref{eq:significance}) for the deviation from the SM $\Mtt$ background of the unnormalised distribution (left) as well as the normalised distributions $\sigma(\Ntwo)$ (centre) and $\sigma(\None)$ (right). The significance is shown as a function of the position of the right edge of a sliding bin with fixed width of $50\GeV$.}
\label{fig:significance}
\end{figure}
The significance of the deviation of signal plus background with respect to pure SM background depends on the type of normalisation, the chosen bin size and the position of the bins. The bin position is especially relevant for BSM contributions with a peak-dip structure: after repositioning, a bin of fixed size can show positive, negative or no deviation from the SM background. Since the minimum size of the $\Mtt$ bins is expected to be around $50\GeV$, i.e. larger than the expected $\Phi$ width, the only way to unambiguously resolve such complicated structure is by using sliding bins. To that end in fig.~\ref{fig:significance} we show the significance:
\begin{equation}
{\rm significance} = {({\rm SM+BSM})_{\rm central} - ({\rm pure~SM})_{\rm central}\over ({\rm pure~SM})_{\rm error}}\,,
\label{eq:significance}
\end{equation}
for a single sliding bin with fixed width of $50\GeV$ as a function of the position of the bin's right edge (i.e. the bin edge that has larger $\Mtt$). 

From fig.~\ref{fig:significance} we conclude that, as expected, the positioning of the bin is very significant for detecting deviations from SM backgrounds. For the unnormalised distribution the significance can be as large as 2. Considering normalised distributions, however, the value of the significance increases significantly. It is as large as 6 for the normalisation $N_{200}$ and reaches 12 for $N_{100}$. Interestingly, the significance of the deviation in the negative direction caused by the interference dip may be large enough to be detectable in normalised distributions.

Finally, we would like to estimate the minimal rate for the process $\sigma(pp\to \Phi\to t\t)$ that could be discriminated from the SM background. In studying this we make the simplifying assumption that the shapes of pure signal and interference remain unchanged and only the overall rate changes. This assumption is roughly consistent with the models considered in ref.~\cite{Hespel:2016qaf}. From fig.~\ref{fig:significance} we observe that for an optimally positioned bin, one can detect with a significance of about 3 a signal with rate $\sigma(pp\to \Phi\to t\t)$ that is as low as $0.55\,{\rm pb}$ for normalisation $N_{200}$ and $\sigma(pp\to \Phi\to t\t)$ that is as low as $0.28\,{\rm pb}$ if normalisation $N_{100}$ is chosen.

\section{Conclusions} 
 
In this work we quantify the possibility for discriminating BSM signals of the type $pp\to \Phi\to t\t$ from the SM $t\t$ background at the LHC 13 TeV by looking for bumps in the $\Mtt$ spectrum. We have found that purposefully normalised $\Mtt$ spectra, computed at NNLO QCD, have small associated theoretical error and can be used as an effective tool for bump-hunting in $t\t$ events. An important property of such normalised distributions is their relatively small sensitivity to the value of the top quark mass. 

To keep the discussion less abstract, as an example, we illustrate our approach by applying it to the case of the current $750\GeV$ di-gamma excess. This excess is important in its own right, given the intense interest into this possible SM deviation. If the $750\GeV$ di-gamma excess is confirmed by forthcoming LHC data (however see sec.~{\it Note Added}), our analysis will provide a workable approach to quantifying the coupling of the resonance $\Phi$ to top quarks. 

Looking beyond the possible $750\GeV$ resonance $\Phi$, our work is designed to be a blueprint into future search strategies for possible resonances decaying to $t\t$ and it can easily be adapted to other kinematic regions. In particular, having high-precision background predictions can be very valuable in designing search strategies in cases where expected bump widths are comparable or smaller than the minimum possible bin size. We expect that our work will offer new insight into designing search strategies in $t\t$ events and will complement and support existing sophisticated bump-hunting statistical techniques and tools \cite{Choudalakis:2011qn}.

\section{Note added}\label{sec:noteadded}

After this paper was submitted for publication, new measurements from the ATLAS and CMS collaborations \cite{ATLAS:2016eeo,Khachatryan:2016yec} were presented at the ICHEP 2016 conference. The new 2016 higher-statistics measurements do not show any indication of a BSM di-gamma signal around $750\GeV$. This implies that the di-gamma excess seen in the 2015 13 TeV LHC data is a statistical fluctuation.

\acknowledgments\vskip4mm
A.M. acknowledges inspiring discussions within the Cambridge SUSY Working Group. The work of M.C. was supported in part by grants of the DFG and BMBF. The work of D.H. and A.M. is supported by the UK Science and Technology Facilities Council [grants ST/L002760/1 and ST/K004883/1].

\appendix\section{Appendix: top quark mass sensitivity of differential distributions}

\begin{figure}[h]
\centering
\hspace{0mm} 
\includegraphics[width=0.46\textwidth]{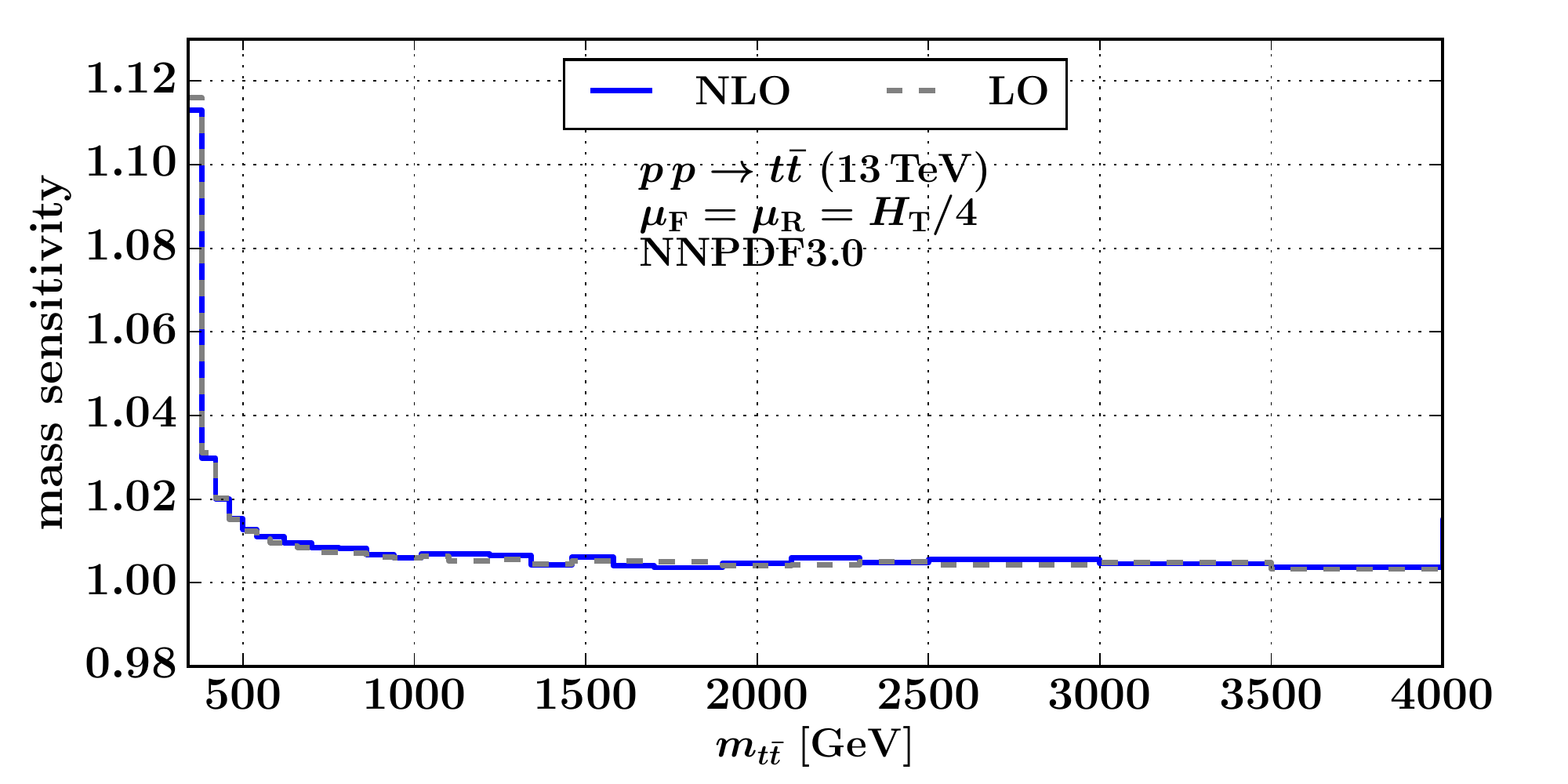}
\includegraphics[width=0.46\textwidth]{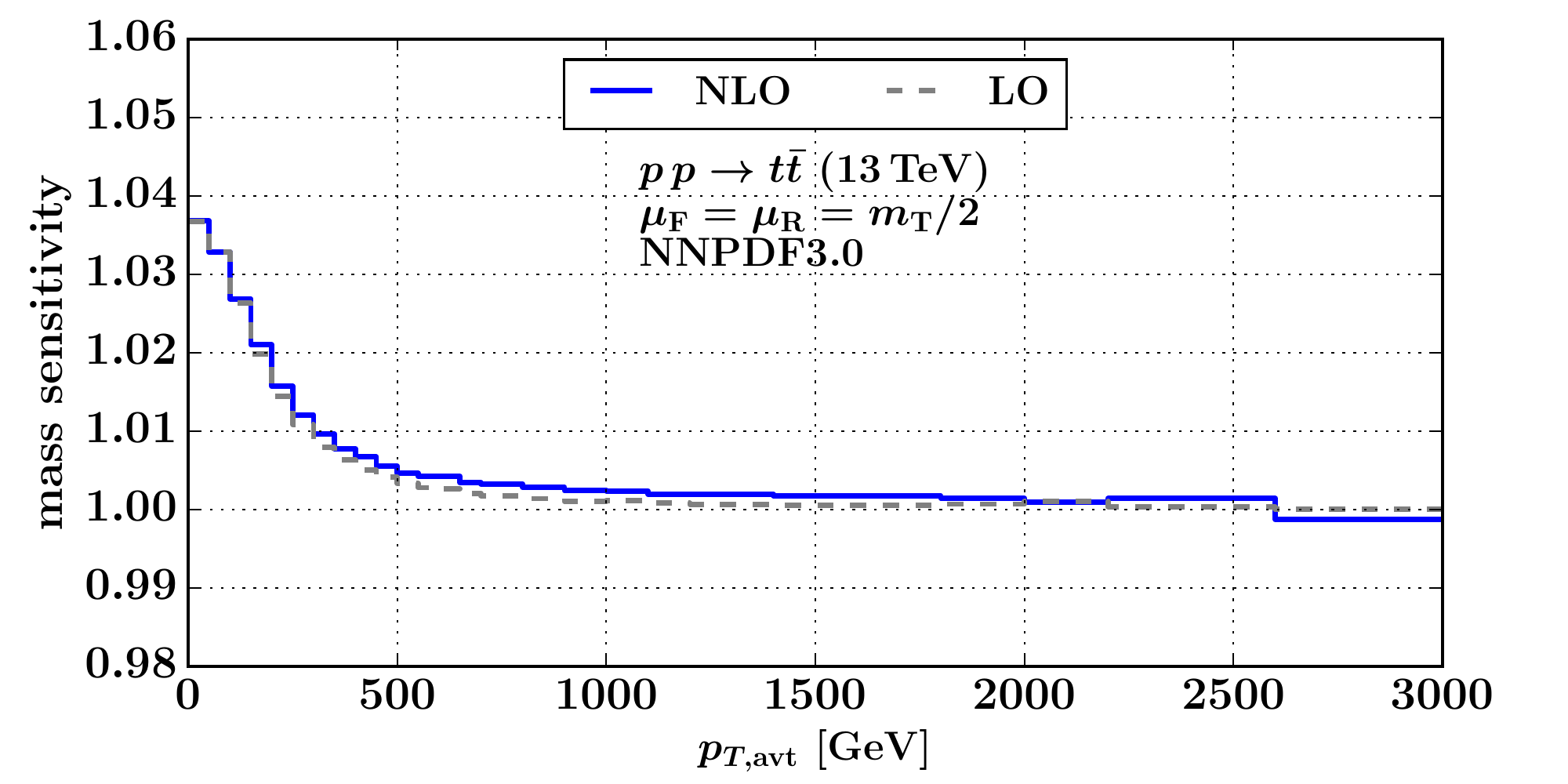}
\includegraphics[width=0.46\textwidth]{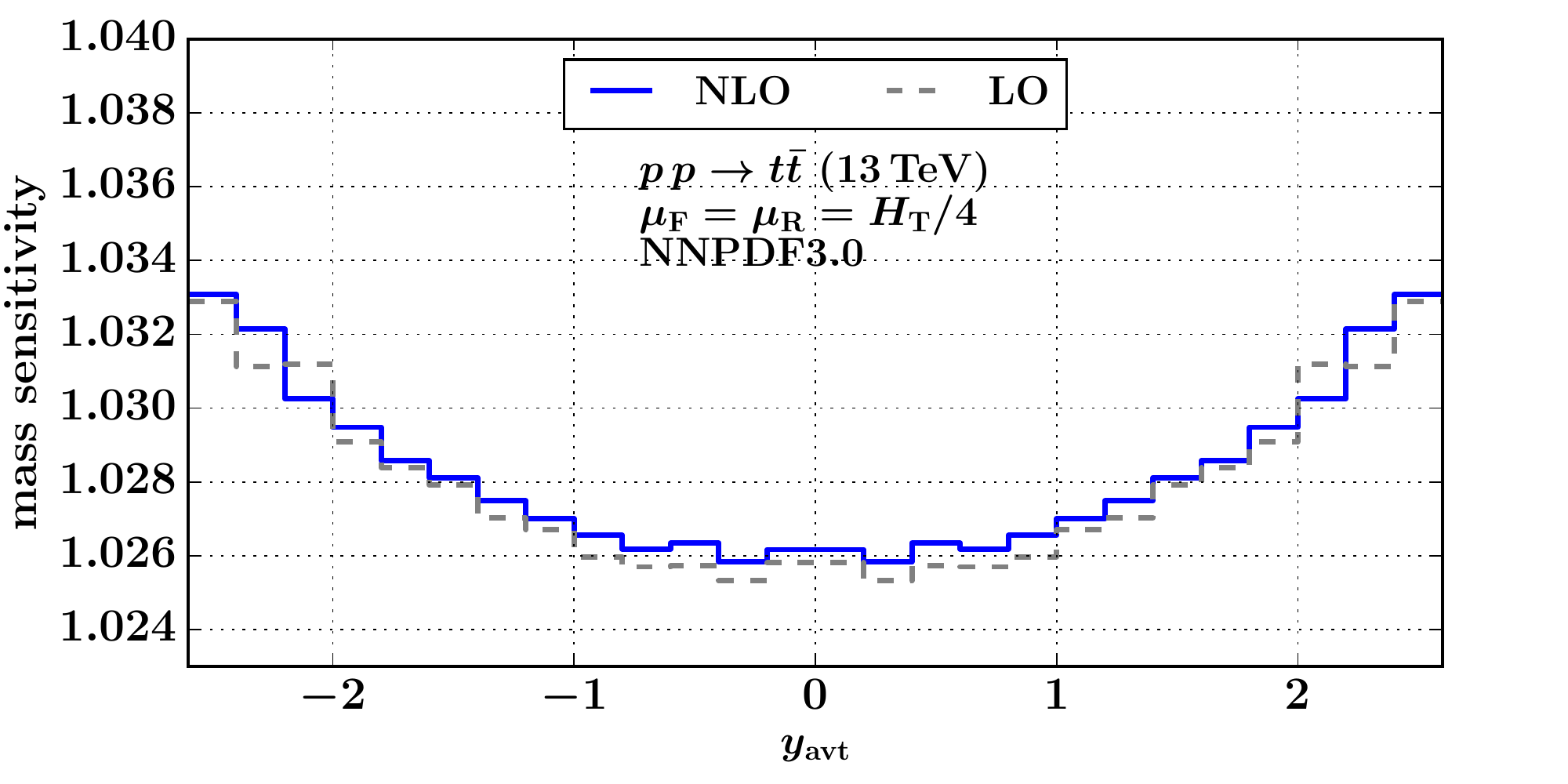}
\includegraphics[width=0.46\textwidth]{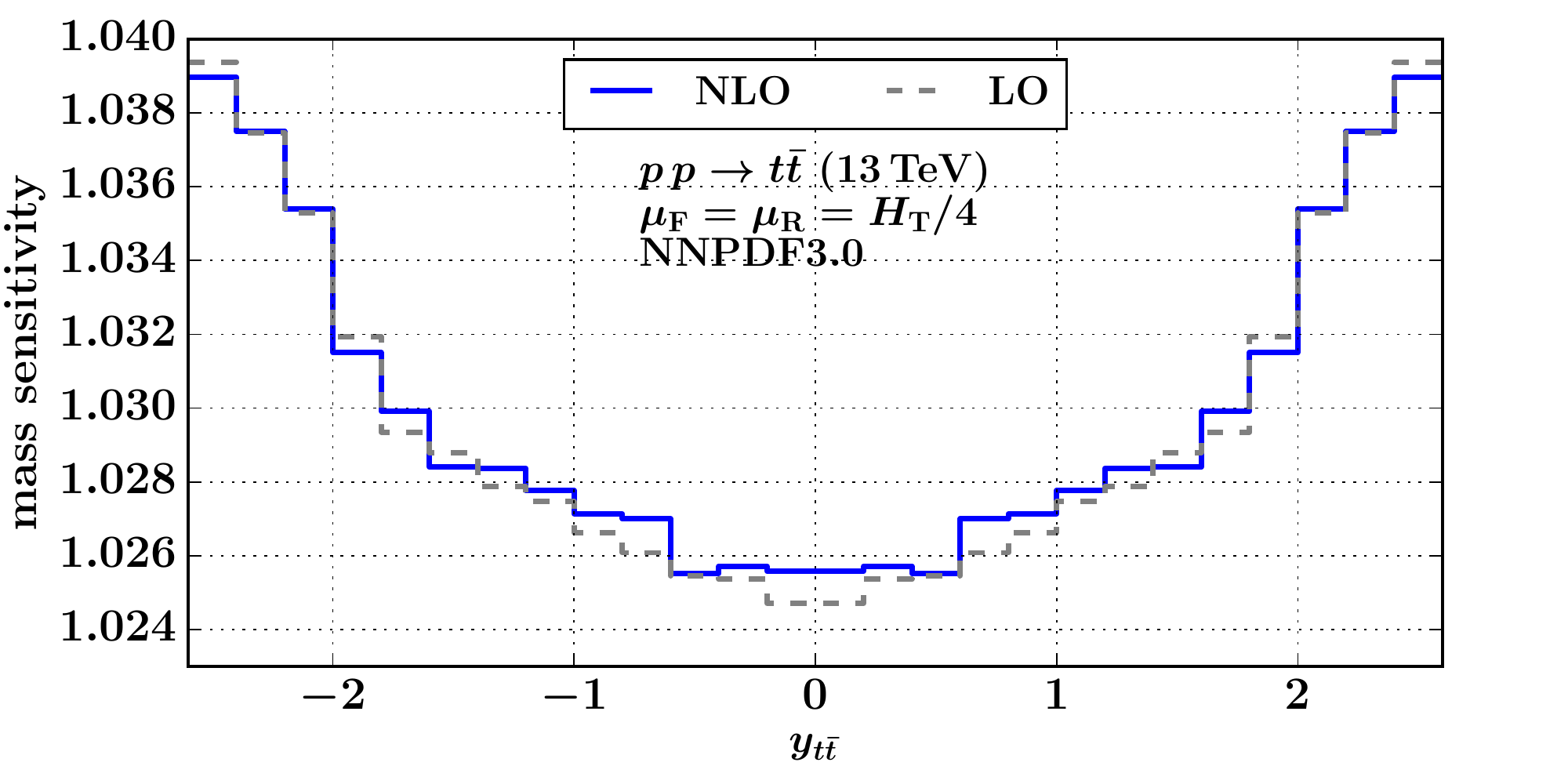}
\caption{Mass sensitivity eq.~(\ref{eq:norm}) of unnormalised distributions: the $t\bar t$ pair's $\Mtt$, $y_{t\bar t}$ and $\PT$, $y_t$ of the average $t/\bar{t}$.}
\label{fig:mass-sensitivities}
\end{figure}

To quantify the $m_t$ sensitivity of the shape of differential distributions we compute the ratio eq.~(\ref{eq:norm}) at LO and NLO for the following four unnormalised distributions: the $t\bar t$ pair's $\Mtt$, $y_{t\bar t}$ and $\PT$, $y_t$ of the average $t/\bar{t}$. The results are shown in fig.~\ref{fig:mass-sensitivities}. 

In fig.~\ref{fig:mass-sensitivities} we observe that the shape of the $\Mtt$ distribution is very sensitive to changes in $m_t$, especially close to threshold. The top $\PT$ distribution is fairly sensitive to $m_t$ close to threshold, although not as much as $\Mtt$. As expected, the mass sensitivity of these unnormalised distributions tends to zero in the limit of large $\Mtt$ or $\PT$. The top and $t\bar t$ rapidities are least sensitive to $m_t$. Their shape sensitivity, however, is rapidly increasing for forward rapidities, especially for $y_{t\bar t}$.

The shape sensitivity of normalised distributions (not shown) is similar to the unnormalised ones in fig.~\ref{fig:mass-sensitivities}. Since the normalisation factor is a kinematics--independent number, its inclusion has the effect of shifting the curves in fig.~\ref{fig:mass-sensitivities} up or down while preserving their shape. We have checked that the mass sensitivity of the normalisation factor, when defined as the total inclusive cross-section, is just under 3\% and changes only by a tiny amount from LO through NNLO. In this calculation we use NNPDF3.0 and always take pdf's and perturbative calculations of matching accuracy.

As we mentioned in the beginning of this work, the tail of the $\Mtt$ and $\PT$ distributions acquires mass sensitivity upon normalisation. This should be anticipated from the results in fig.~\ref{fig:mass-sensitivities} since the tails of the absolute $\Mtt$ and $\PT$  distributions are not $m_t$ sensitive while the normalisation factor is.


\begin{thebibliography}{99}

\bibitem{Spano:2013nca} 
  F.~Spano,
  EPJ Web Conf.\  {\bf 55}, 03002 (2013).

\bibitem{ATLAS750note} 
  The ATLAS collaboration,
  ATLAS-CONF-2015-081.

\bibitem{CMS:2015dxe} 
 The CMS collaboration,
  CMS-PAS-EXO-15-004.

\bibitem{Aaboud:2016tru} 
  M.~Aaboud {\it et al.} [ATLAS Collaboration],
  arXiv:1606.03833 [hep-ex].
  
\bibitem{Khachatryan:2016hje} 
  V.~Khachatryan {\it et al.} [CMS Collaboration],
  arXiv:1606.04093 [hep-ex].  

\bibitem{Strumia:2016wys} 
  A.~Strumia,
  arXiv:1605.09401 [hep-ph].

\bibitem{Aad:2015fna} 
  G.~Aad {\it et al.} [ATLAS Collaboration],
  JHEP {\bf 1508}, 148 (2015)
  [arXiv:1505.07018 [hep-ex]].

\bibitem{Aad:2013nca} 
  G.~Aad {\it et al.} [ATLAS Collaboration],
  Phys.\ Rev.\ D {\bf 88}, no. 1, 012004 (2013)
  [arXiv:1305.2756 [hep-ex]].

\bibitem{Aad:2012dpa} 
  G.~Aad {\it et al.} [ATLAS Collaboration],
  JHEP {\bf 1209}, 041 (2012)
  [arXiv:1207.2409 [hep-ex]].

\bibitem{Hespel:2016qaf} 
  B.~Hespel, F.~Maltoni and E.~Vryonidou,
  arXiv:1606.04149 [hep-ph].

\bibitem{Djouadi:2016ack} 
  A.~Djouadi, J.~Ellis and J.~Quevillon,
  arXiv:1605.00542 [hep-ph].

\bibitem{Bernreuther:2015fts} 
  W.~Bernreuther, P.~Galler, C.~Mellein, Z.~G.~Si and P.~Uwer,
  Phys.\ Rev.\ D {\bf 93}, no. 3, 034032 (2016)
  [arXiv:1511.05584 [hep-ph]].

\bibitem{Craig:2015jba} 
  N.~Craig, F.~D'Eramo, P.~Draper, S.~Thomas and H.~Zhang,
  JHEP {\bf 1506}, 137 (2015)
  [arXiv:1504.04630 [hep-ph]].
  
\bibitem{Gori:2016zto} 
  S.~Gori, I.~W.~Kim, N.~R.~Shah and K.~M.~Zurek,
  Phys.\ Rev.\ D {\bf 93}, no. 7, 075038 (2016)
  [arXiv:1602.02782 [hep-ph]].

\bibitem{Czakon:2016dgf} 
  M.~Czakon, D.~Heymes and A.~Mitov,
  arXiv:1606.03350 [hep-ph].
  
\bibitem{Czakon:2015owf} 
  M.~Czakon, D.~Heymes and A.~Mitov,
  Phys.\ Rev.\ Lett.\  {\bf 116}, no. 8, 082003 (2016)
  [arXiv:1511.00549 [hep-ph]].  

\bibitem{Ball:2014uwa} 
  R.~D.~Ball {\it et al.} [NNPDF Collaboration],
  JHEP {\bf 1504}, 040 (2015)
  [arXiv:1410.8849 [hep-ph]].

\bibitem{Cacciari:2008zb} 
  M.~Cacciari, S.~Frixione, M.~L.~Mangano, P.~Nason and G.~Ridolfi,
  JHEP {\bf 0809}, 127 (2008)
  [arXiv:0804.2800 [hep-ph]].

\bibitem{Czakon:2016ckf} 
  M.~Czakon, P.~Fiedler, D.~Heymes and A.~Mitov,
  JHEP {\bf 1605}, 034 (2016)
  [arXiv:1601.05375 [hep-ph]].

\bibitem{Czakon:2016olj} 
  M.~Czakon, N.~P.~Hartland, A.~Mitov, E.~R.~Nocera and J.~Rojo,
  arXiv:1611.08609 [hep-ph].

\bibitem{Pagani:2016caq} 
  D.~Pagani, I.~Tsinikos and M.~Zaro,
  arXiv:1606.01915 [hep-ph].

\bibitem{NNLO-EW} 
  M.~Czakon, D.~Heymes, A.~Mitov, D.~Pagani, I.~Tsinikos and M.~Zaro,
  To appear.

\bibitem{ATLAS:2014wva} 
  [ATLAS and CDF and CMS and D0 Collaborations],
  arXiv:1403.4427 [hep-ex].

\bibitem{CMS:2014ima} 
  CMS Collaboration [CMS Collaboration],
  CMS-PAS-TOP-14-001.

\bibitem{Abazov:2014dpa} 
  V.~M.~Abazov {\it et al.} [D0 Collaboration],
  Phys.\ Rev.\ Lett.\  {\bf 113}, 032002 (2014)
  [arXiv:1405.1756 [hep-ex]].  

\bibitem{Stieger:2016vgj} 
  B.~Stieger,
  arXiv:1606.02482 [hep-ex].

\bibitem{Frixione:2014ala} 
  S.~Frixione and A.~Mitov,
  JHEP {\bf 1409}, 012 (2014)
  [arXiv:1407.2763 [hep-ph]].

\bibitem{Harland-Lang:2014zoa} 
  L.~A.~Harland-Lang, A.~D.~Martin, P.~Motylinski and R.~S.~Thorne,
  Eur.\ Phys.\ J.\ C {\bf 75}, no. 5, 204 (2015)
  [arXiv:1412.3989 [hep-ph]].

\bibitem{Czakon:2014xsa} 
  M.~Czakon, P.~Fiedler and A.~Mitov,
  Phys.\ Rev.\ Lett.\  {\bf 115}, no. 5, 052001 (2015)
  [arXiv:1411.3007 [hep-ph]].

\bibitem{Choudalakis:2011qn} 
  G.~Choudalakis,
  arXiv:1101.0390 [physics.data-an].                  

\bibitem{ATLAS:2016eeo} 
  The ATLAS collaboration [ATLAS Collaboration],
  ATLAS-CONF-2016-059.

\bibitem{Khachatryan:2016yec} 
  V.~Khachatryan {\it et al.} [CMS Collaboration],
  arXiv:1609.02507 [hep-ex].

\end{thebibliography}
\end{document}